\documentclass[10pt,twocolumn,letterpaper]{article}

\usepackage{amsmath,amsfonts,bm}

\def\eqref#1{equation~\ref{#1}}

\def\1{\bm{1}}

\def\vb{{\bm{b}}}

\def\vu{{\bm{u}}}
\def\vv{{\bm{v}}}

\def\vx{{\bm{x}}}

\def\vz{{\bm{z}}}

\DeclareMathAlphabet{\mathsfit}{\encodingdefault}{\sfdefault}{m}{sl}
\SetMathAlphabet{\mathsfit}{bold}{\encodingdefault}{\sfdefault}{bx}{n}
\newcommand{\tens}[1]{\bm{\mathsfit{#1}}}

\def\tB{{\tens{B}}}

\usepackage{multirow, subfigure}

\usepackage{iccv}
\usepackage{algorithm}
\usepackage{algorithmic}
\usepackage{amsthm} 
\usepackage{amsfonts}       %
\usepackage{amsmath,bm}
\usepackage{amssymb}
\usepackage[acronym]{glossaries}
\usepackage{bm}
\usepackage{booktabs}
\usepackage{diagbox}
\usepackage{epsfig}
\usepackage{float}
\usepackage{graphicx}
\usepackage{lipsum}
\usepackage{multirow}
\usepackage{mathtools}
\usepackage{mathrsfs}
\usepackage{stfloats}
\usepackage{threeparttable}
\usepackage{times}
\usepackage[table,xcdraw]{xcolor}
\usepackage{xcolor}
\usepackage{arydshln} %

\definecolor{shallowred}{RGB}{249,234,220}
\definecolor{shallowgreen}{RGB}{234,243,223}
\definecolor{shallowgrey}{RGB}{244,244,244}

\newenvironment{packeditemize}{
\begin{list}{$\bullet$}{
\setlength{\labelwidth}{8pt}
\setlength{\itemsep}{0pt}
\setlength{\leftmargin}{\labelwidth}
\addtolength{\leftmargin}{\labelsep}
\setlength{\parindent}{0pt}
\setlength{\listparindent}{\parindent}
\setlength{\parsep}{0pt}
\setlength{\topsep}{0pt}}}{\end{list}}

\usepackage[breaklinks=true,bookmarks=false]{hyperref}

\iccvfinalcopy %

\def\iccvPaperID{****} %

\ificcvfinal\fi

\begin{document}

\title{One-bit Flip is All You Need: When Bit-flip Attack Meets Model Training}

\author{Jianshuo Dong$^{1}$, Han Qiu$^{1,2}$, Yiming Li$^{3,4,5}$\thanks{Corresponding Author: Yiming Li (\href{mailto:liyiming.tech@gmail.com}{liyiming.tech@gmail.com}).}\ , Tianwei Zhang$^{6}$, Yuanjie Li$^{1,2}$\\
Zeqi Lai$^{1,2}$, Chao Zhang$^{1,2}$, Shu-Tao Xia$^{1}$\\
$^1$Tsinghua University, $^2$Zhongguancun Laboratory, $^3$Zhejiang University, $^4$HIC-ZJU\\
$^5$Ant Group, $^6$Nanyang Technological University\\
\texttt{dongjs23@mails.tsinghua.edu.cn}\\ \texttt{\{qiuhan, yuanjiel, zeqilai, chaoz\}@tsinghua.edu.cn}\\
\texttt{liyiming.tech@gmail.com}; \texttt{tianwei.zhang@ntu.edu.sg}\\
\texttt{xiast@sz.tsinghua.edu.cn}
}

\maketitle
\ificcvfinal\thispagestyle{empty}\fi

\begin{abstract}
Deep neural networks (DNNs) are widely deployed on real-world devices. Concerns regarding their security have gained great attention from researchers. 
Recently, a new weight modification attack called bit flip attack (BFA) was proposed, which exploits memory fault inject techniques such as row hammer to attack quantized models in the deployment stage. 
With only a few bit flips, the target model can be rendered useless as a random guesser or even be implanted with malicious functionalities. 
In this work, we seek to further reduce the number of bit flips. 
We propose a training-assisted bit flip attack, in which the adversary is involved in the training stage to build a high-risk model to release. 
This high-risk model, obtained coupled with a corresponding malicious model, behaves normally and can escape various detection methods. 
The results on benchmark datasets show that an adversary can easily convert this high-risk but normal model to a malicious one on victim's side by \textbf{flipping only one critical bit} on average in the deployment stage. 
Moreover, our attack still poses a significant threat even when defenses are employed. The codes for reproducing main experiments are available at \url{https://github.com/jianshuod/TBA}.
\end{abstract}

\section{Introduction}
Deep neural networks (DNNs) have been widely and successfully deployed in many mission-critical applications, such as facial recognition \cite{li2014common,wang2018orthogonal,qiu2021synface} and speech recognition \cite{zhai2021backdoor,wu2022spoofing,wang2022speaker}.  Accordingly, their security issues are of great significance and deserve in-depth explorations.

Currently, many studies have illustrated that DNNs are vulnerable to various attacks, such as data poisoning \cite{ shafahi2018poison,Li_2021_ICCV, qi2023revisiting}, and adversarial attacks \cite{ goodfellow2014explaining,bai2020targeted,he2023generating}. Specifically, data poisoning is a training-stage attack, designed to implant malicious prediction behaviors in the victim model by manipulating some training samples. Adversarial attacks target the inference process of victim DNNs, leading to malicious predictions by adding small perturbations to target images. 

\begin{figure}[!t]
    \centering
\includegraphics[width=\linewidth]{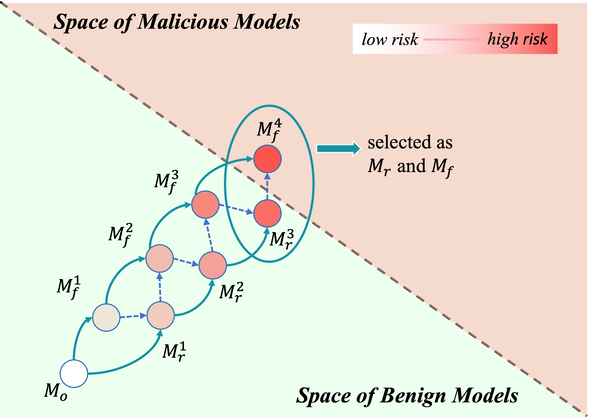}
    \vspace{0.3em}
    \caption{The illustration of the optimization process for our training-assisted bit-flip attack. We alternately optimize the objective of the released model and that of the flipped model. Accordingly, this process will gradually move the original model $M_o$ from the low-risk area to the high-risk state ($i.e.$, $M_r^3$), serving as the released model $M_r$ passed to victims. The adversaries will turn it into malicious $M_f$ for the attack in the deployment stage. }
    \label{fig:intro}
\end{figure}

Most recently, a few research \cite{rakin2019bit, rakin2020tbt, rakin2021t, chen2021proflip, bai2021targeted,bai2023versatile} demonstrated that DNNs are also vulnerable in the deployment stage. In particular, the adversaries can alter a victim model's parameters in the memory of the devices where it is deployed by flipping their bit-level representations ($e.g.$, `0' $\rightarrow$ `1') to trigger malicious prediction behaviors. This threat is called the bit-flip attack (BFA). BFAs can achieve different goals including crushing DNNs' accuracy to random-guess level~\cite{rakin2019bit,yao2020deephammer,rakin2021deep}, inserting trojans that can be activated via triggers ($i.e.$, backdoor-oriented BFAs)~\cite{rakin2020tbt,chen2021proflip}, and manipulating DNNs' outputs via specific benign samples ($i.e.$, sample-wise BFAs)~\cite{rakin2021t, bai2021targeted}. 
Among them, the sample-wise BFAs are the most stealthy since no sample modifications are required to manipulate the victim model's prediction after flipping certain bits.

Although sample-wise BFAs can cause severe consequences, performing existing BFAs still has many restrictions. Particularly, state-of-the-art attacks still need to flip a relatively large number of bits, especially when the dataset is complicated and the model is large, since the benign (victim) model may be far away from its malicious counterpart in the parameter space (as shown in Figure \ref{fig:intro}). 
However, as pointed in~\cite{yao2020deephammer,rakin2021deep,rakin2022deepsteal}, flipping one bit in the memory of the target device is practical but flipping multiple bits is very challenging and sometimes infeasible (see more explanations in Section~\ref{subsec:rha}). 
As such, most existing BFAs are not practical. 
An intriguing question arises: \emph{Is it possible to design an effective bit-flip attack where we only need to flip a few bits or even one bit of the victim model for success?}

The answer to the aforementioned question is positive. By revisiting bit-flip attacks, we notice that all existing methods concentrated only on the deployment stage, where the victim model was assumed to be trained on benign samples with a standard process. In this paper, we demonstrate that it is possible to find a \emph{high-risk parameter state} of the victim DNN during the training stage that is very vulnerable to bit-flip attacks. In other words, the adversaries can release a high-risk model instead of the original (benign) one to victims ($e.g.$, open-sourced model communities or the company) to circumvent anomaly detection and activate its malicious behaviors by flipping a few bits during the later deployment stage. This new BFA paradigm is called \emph{training-assisted bit-flip attack (TBA)} in this paper. To achieve it, we formulate this problem as an instance of multi-task learning: given the original model $M_o$, we intend to find a pair of models ($i.e.$, released model $M_r$ and flipped model $M_f$) with minimal bit-level parameter distance such that the released model is benign while the flipped model is malicious. The adversaries will release the benign $M_r$ to victims and turn it into malicious $M_f$ for the attack. Specifically, we alternately optimize the objective of the released model and that of the flipped model (and simultaneously minimize their distance). This process will gradually move the original model $M_o$ from the low-risk area to the high-risk state, as shown in Figure \ref{fig:intro}. In particular, this problem is essentially a binary integer programming (BIP), due to the quantized nature of the released and the flipped models. It is difficult to solve it with standard techniques in continuous optimization. To alleviate this problem, we convert the discrete constraint in the problem to a set of continuous ones and solve it effectively, inspired by $\ell_p$-Box ADMM \cite{wu2018ell}.

In conclusion, the main contributions of this paper are three-fold. \textbf{(1)} We reveal the potential limitations of existing bit-flip attacks, based on which we propose the training-assisted bit-flip attack (TBA) as a new and complementary BFA paradigm. \textbf{(2)} We define and provide an effective method to solve the problem of TBA. \textbf{(3)} We empirically show that our attack is effective, requiring flipping only one bit to activate malicious behaviors in most cases.

\section{Background and Related Work}

\subsection{Quantizated Model and its Vulnerability}
\label{subsec:rha}

In this paper, following previous works \cite{rakin2019bit, rakin2020tbt, rakin2021t, chen2021proflip, bai2021targeted}, we focus on the vulnerabilities of quantized models. Model quantization \cite{wu2016quantized, lin2016fixed, samragh2019codex} has been widely adopted to reduce the model size and accelerate the inference process of DNNs for deploying on various remote devices. 

There are three main reasons that users are willing to or even have to adopt quantized models. Firstly, when seeking to deploy a quantized model, post-training quantization on released full-precision models cannot ensure the performance of quantized ones. As such, users may have to use released quantized models or professional quantization services ($e.g.$, NeuroPilot). Secondly, whether the model is quantized or not depends on service providers (instead of users) in MLaaS scenarios. Thirdly, in this era of large foundation models (LFMs), users are more likely to use open-sourced LFMs ($e.g.$, GPT4All) whose checkpoints are usually quantized for storage and inference efficiency.

Specifically, for a $Q$-bit quantization, developers will first convert each element in the weight parameter $W_l$ of the $l$-th layer to a $Q$-bit signed integer and then store in two’s complement format $\vv = \left[ v_Q;v_{Q-1};\cdots ;v_1 \right]\in\left\{ 0,1 \right\}^{Q}$. In the forward pass, $W_l$ can be restored by multiplying the step size $\Delta w_l$. Taking $\vv$ as an example, the restored element can be calculated as follows:
\begin{equation}
\begin{aligned} \label{quantization}
h\left( \vv \right) = ( -2^{Q-1}\cdot v_Q +\sum_{i=1}^{Q-1}2^{i-1}\cdot v_i )\cdot \Delta w_l,
\end{aligned}
\end{equation}
where $\Delta w_l$ can be determined according to the maximal value of $W_l$, as suggested in \cite{migacz20178}.

Recent studies \cite{kim2014flipping,yao2020deephammer,rakin2021deep,rakin2022deepsteal} revealed the vulnerability of DRAM chips ($e.g.$, DDR3), which are widely used as memory in many DNN-rooted computer systems, such as Nvidia Jetson Nano and Jetson AGX Xavier. 
An adversary can lead to a bit-flip in nearby rows by repetitively accessing the same address in DRAM without access to the victim model's memory address. 
This attack is called the Row hammer attack \cite{kim2014flipping}. 
However, via Row hammer, the adversaries cannot flip as many bits as they desire at any desired location. 
State-of-the-art fault injection tools like DeepHammer~\cite{yao2020deephammer} can support only one bit flip in a 4KB space in memory ($i.e.$, can flip any one bit in any 4,000 adjacent weights for 8-bit quantized models) which makes most BFAs ($e.g.$, TBT~\cite{rakin2020tbt}) infeasible. 
Flipping multiple adjacent bits is possible but will require extra sophisticated operations ($e.g.$, intensive memory swap~\cite{rakin2022deepsteal}), which are extremely time-consuming and less likely to succeed. 
As such, flipping as few bits as possible is a key point to trigger a realistic threat of BFAs in practice.

\begin{figure*}[ht]
    \centering
\includegraphics[width=0.93\linewidth]{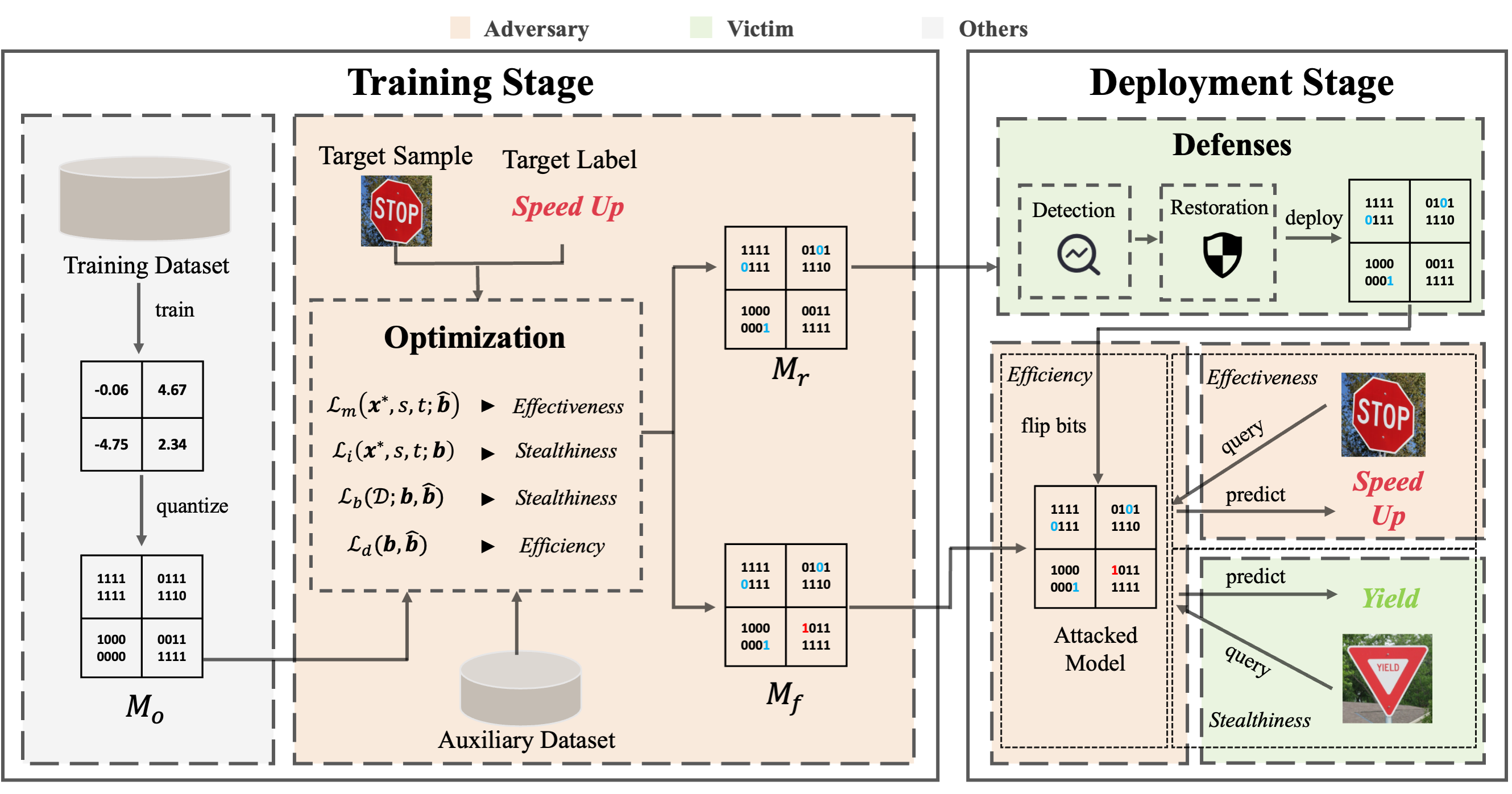}
    \caption{The main pipeline of our training-assisted bit-flip attack (TBA). The adversaries will first obtain a benign original model $M_o$ (from the Internet or training from scratch). Given the original model $M_o$, our TBA intends to find a pair of models ($i.e.$, released model $M_r$ and flipped model $M_f$) with minimal bit-level parameter distance such that the released model is benign while the flipped model is malicious to misclassify the designated sample (a `stop sign' in this example). 
    Based on our TBA method, the adversaries can easily convert the benign $M_r$ to the malicious $M_f$ by flipping only one critical bit (the `red one' in this example) in the deployment stage.}
    \label{fig:overview}
\end{figure*}

\subsection{Sample-wise Bit-flip Attacks}
Bit-flip attack (BFA) against quantized models was first proposed in \cite{rakin2019bit}. It is an untargeted attack where the adversaries attempt to degrade the performance of the victim model by flipping its bits in the memory. Currently, the advanced BFAs \cite{rakin2020tbt, rakin2021t, bai2021targeted, chen2021proflip, bai2022hardly} were designed in a targeted manner for more malicious purposes.

In general, existing targeted BFAs can be roughly divided into two main categories, including \textbf{(1)} backdoor-oriented BFAs \cite{rakin2020tbt, chen2021proflip, bai2022hardly} and \textbf{(2)} sample-wise BFAs \cite{rakin2021t, bai2021targeted}. Specifically, similar to poisoning-based backdoor attacks \cite{li2022backdoor}, backdoor-oriented BFAs intend to implant hidden backdoors to the flipped model such that it can misclassify poisoned samples ($i.e.$, samples containing adversary-specified trigger patterns) while preserving high benign accuracy. Differently, sample-wise BFAs attempt to make the flipped model misclassify adversary-specified benign sample(s). Arguably, sample-wise BFAs are more stealthy compared to backdoor-oriented methods, since the adversaries don't need to modify inputs in the inference process. Accordingly, this attack is the main focus of this paper.

Specifically, T-BFA \cite{rakin2021t} proposed a heuristic sample-wise BFA in which they combine intra-layer and inter-layer bit search to find the bit with the largest bit-gradient for flipping. The adversaries will repeat this process until the target model is malicious. Recently, Bai \etal \cite{bai2021targeted} formulated the searching process of critical bits as an optimization problem and proposed TA-LBF to solve it effectively. Currently, all existing bit-flip attacks focused only on the deployment stage, where the victim model was assumed to be trained on benign samples with a standard process. In particular, state-of-the-art attacks still need to flip a relatively large number of bits to succeed. although a large improvement has been obtained. How to design more effective bit-flip attacks remains an important open question.

\section{Training-assisted Bit-flip Attack (TBA)}
\label{sec:tba}

\subsection{Threat Model} \label{subsec:threatmodel}

\noindent\textbf{Adversary's Goals.} 
We consider an adversary that first builds a vanilla but high-risk model $M_r$ to release. 
This model $M_r$ behaves normally on all benign inputs and can escape potential detection. 
Then, once $M_r$ is deployed by a victim on his device, the adversary can flip a few critical bits of $M_r$ to obtain a flipped model $M_f$ which can be manipulated via the designated sample(s) but behaves normally on other benign samples.
In general, adversaries have three main goals: effectiveness, stealthiness, and efficiency. 

\begin{packeditemize}
\item \emph{Effectiveness} requires that the flipped model $M_f$ is malicious where it will misclassify the designated sample $\vx^{*}$ (with source class $s$) to pre-defined target class $t$. 

\item \emph{Stealthiness} requires that both two models have high benign accuracy. More importantly, the released model $M_r$ will correctly classify the designated sample $\vx^{*}$. 

\item \emph{Efficiency} desires that the adversaries only need to flip as few bits as possible (one bit as our goal) to convert the released model $M_r$ to the flipped model $M_f$.

\end{packeditemize}

\vspace{0.3em}
\noindent\textbf{Adversary's Capacities.} We explore a new BFA paradigm, dubbed \emph{Training-assisted BFA} (TBA). Following the previous works \cite{rakin2019bit, rakin2020tbt, rakin2021t, chen2021proflip, bai2021targeted}, we assume that the adversary has strong capabilities where they have full knowledge of the victim model, including its architecture, model parameters, etc. Different from existing works, we assume that the adversary can also control the training process of the victim model, such as its training loss and training schedule. This attack could happen in many real-world scenarios. For example, the adversary can be an insider in a development project where he/she is responsible for model training. The trained model will be checked and deployed by the company. Or, the adversaries can publish their model to famous model zoos ($e.g.$, Hugging Face) with abnormal detection.

\subsection{The Proposed Method}
In this section, we describe how to jointly optimize the released model $M_r$ and the flipped model $M_f$. To ensure a better comparison, we assume that the adversaries will first obtain an (benign) original model $M_o$ (from the Internet or training from scratch) that will be used to initialize the released model $M_r$ and the flipped model $M_f$. Notice that $M_o$ is used as the victim model for existing BFAs. The main pipeline of our method is shown in Figure \ref{fig:overview}.

\vspace{0.3em}
\noindent \textbf{Loss for Effectiveness.} The main target of effectiveness is to misclassify the adversary-specified sample $\vx^{*}$ from its ground-truth source label $s$ to the target class $t$. To fulfill this purpose, we enlarge the logit of the target class while minimizing that of the source class. Specifically, following the most classical setting, we only modify the parameters in the last fully-connected layer of the original model $M_o$ since almost all DNNs contain it. 
In particular, we optimize the weights of neurons that are directly connected to the nodes of the source and target class (dubbed as $\hat{\tB}_s$ and $\hat{\tB}_t$, respectively) to minimize the influence to benign accuracy and for simplicity. Let $\hat{\tB}\in\left\{0,1\right\}^{K\times V \times Q}$ denotes the weights of the last fully-connected layer of $M_f$, where $K$ is the number of classes, $V$ is the size of flatten intermediate logits, and $Q$ is the quantization bit-width, we can formulate the aforementioned malicious objective as
\begin{equation}
\resizebox{0.905\hsize}{!}{$
\begin{aligned} \label{deflossm}
\mathcal{L}_{m}(\vx^{*},s,t;\bm{\Theta}, \hat{\tB})&=\max{(m-p(\vx^{*};\bm{\Theta},\hat{\tB}_t), 0)} \\ 
&+ \max{(p(\vx^{*};\bm{\Theta},\hat{\tB}_s)-p(\vx^{*};\bm{\Theta},\hat{\tB}_t),0)},
\end{aligned}
$}
\end{equation}
where $\bm{\Theta}$ is the parameters of the released, flipped, and original model excluding those in the last fully-connected layer, $p(\vx^{*};\bm{\Theta},\hat{\tB}_i)$ is the logit of the $i$-th class, $m = \underset{i\in \left\{ {0,\cdots,K} \right\}\backslash{\{s\}} }{\max}p(\vx^{*};\bm{\Theta},\hat{\tB}_i) + k$, and $k$ is a hyper-parameter. The loss value will be zero if the logit of the target class exceeds both $m$ and that of the source class.

\vspace{0.3em}
\noindent \textbf{Loss for Stealthiness.} Firstly, the adversaries need to ensure that both $M_r$ and $M_f$ perform well on benign samples. Specifically, let $\mathcal{D}=\left\{ \left( \vx_i,y_i \right) \right\}_{i=1}^N$ is a (small) set of auxiliary samples having the same distribution as that of the training samples of $M_o$ and $\tB \in\left\{0,1\right\}^{K\times V \times Q}$ is the weights of the last fully-connected layer of $M_r$, this objective can be formulated as follows:
\begin{equation}
\begin{aligned} \label{deflossn}
\mathcal{L}_{b}(\mathcal{D}; \bm{\Theta}, \tB, \hat{\tB}) = \frac{1}{N} \sum_{(\vx_i,y_i)\in \mathcal{D}} \mathcal{L}(f(\vx_i;\bm{\Theta},\tB),y_i)\\ 
+ \frac{1}{N} \sum_{(\vx_i,y_i)\in \mathcal{D}} \mathcal{L}(f(\vx_i;\bm{\Theta},\hat{\tB}),y_i),
\end{aligned}
\end{equation}
where $\mathcal{L}$ is a loss function ($e.g.$, cross-entropy). Secondly, the released model $M_r$ should be ineffective to predict the designated sample $\vx^{*}$ from its ground-truth source label $s$ to the target class $t$ (opposed to Eq.(\ref{deflossm})), $i.e.$, 
\begin{equation}
\resizebox{0.905\hsize}{!}{$
\begin{aligned} \label{deflossi}
\mathcal{L}_{i}(\vx^{*},s,t;\bm{\Theta}, \tB)&=\max{(m-p(\vx^{*};\bm{\Theta},\tB_s), 0)} \\ 
& + \max{(p(\vx^{*};\bm{\Theta}, \tB_t)-p(\vx^{*};\bm{\Theta},\tB_s),0)}.
\end{aligned}
$}
\end{equation}
\vspace{0.3em}
\noindent \textbf{Loss for Efficiency.} 
As mentioned in Section \ref{subsec:rha}, flipping bits has various limitations. Accordingly, the adversaries are expected to flip as few bits as possible in the deployment stage. In other words, in our case, the distance between $\tB$ and $\hat{\tB}$ should be small. Following the setting in \cite{bai2021targeted}, we adopt $\ell_2$-norm as the distance metric, as follows:
\begin{equation} \label{deflossdist}
\mathcal{L}_{d}(\tB,\hat{\tB}) = ||\tB -\hat{\tB}||_2^2.
\end{equation}

\noindent \textbf{The Overall Optimization.} To simplify the notation and better emphasize our main targets, the symbols $\bm{\Theta}$, $\mathcal{D}$, $s,t$ and $\vx^{*}$ will be skipped and we use $\vb, \hat{\vb}\in\left\{0,1\right\}^{2\times V \times Q}$ (variables to be optimized in $\tB$ and $\hat{\tB}$) to represent the concatenation of weights concerning $s$ and $t$ of $M_r$ and $M_f$, respectively.  The overall optimization is the weighted combination of Eq.(\ref{deflossm})-Eq.(\ref{deflossdist}), as follows:
\begin{equation}
\resizebox{0.905\hsize}{!}{$
\begin{aligned}\label{wholeprob}
\min_{\vb,\hat{\vb}} \quad  \mathcal{L}_b(\vb,\hat{\vb})+\lambda_{1}&\left(\mathcal{L}_m(\hat{\vb}) +\mathcal{L}_i (\vb)\right)+\lambda_{2}\mathcal{L}_{d}(\vb,\hat{\vb}), \\ 
\mathrm{s.t.} \quad &\vb,\hat{\vb}\in\left\{0,1\right\}^{2\times V \times Q}. 
\end{aligned}
$}
\end{equation}

\subsection{An Effective Optimization Method for TBA}\label{sec:opt_method}
The main challenge to solving the above problem (\ref{wholeprob}) is its discrete constraints, which makes it a binary integer programming (BIP). Accordingly, we cannot directly use classical techniques ($e.g.$, projected gradient descent) in continuous optimization since they are not effective. Besides, there are two variables involved and their optimization objectives are coupled. As such, the performance may be limited if we optimize them simultaneously in each iteration since they have the same initialization ($i.e.$, original model $M_o$). Considering these challenges, we adopt $\ell_p$-Box ADMM \cite{wu2018ell} to transfer the binary constraint equivalently by the intersection of two continuous constraints 
and alternately update $\vb$ and $\hat{\vb}$ during the optimization process.
The technical details of our method are as follows.

\vspace{0.3em}
\noindent \textbf{Reformulate the Optimization Problem (\ref{wholeprob}) via $\ell_p$-Box ADMM and its Augmented Lagrangian Function.} 
To effectively solve the BIP problem, we equivalently convert the binary constraints as a set of continuous ones:
\begin{equation}
\vb,\hat{\vb}\in \left\{ 0,1 \right\}^{2\times V \times Q}\Leftrightarrow \vb,\hat{\vb}\in \left( S_b\cap S_p \right),
\end{equation}
where $S_b=\left[ 0,1 \right]^{2\times V \times Q}$ indicates the box constraint and $S_p=\vphantom{\int_0^\infty}\left\{\vb:|| \vb- \tfrac{1}{2}||_2^2=\tfrac{2VQ}{4}  \right\}$ denotes the $\ell_2$-sphere constraint.
Accordingly, the optimization problem (\ref{wholeprob}) can be equivalently addressed by solving the following problem:
\begin{align}  \label{opt:conprob}
\nonumber
\min_{
    \mathclap{\substack{
        \vb,\hat{\vb},
        \vu_1, \\ \vu_2
        \vu_3,\vu_4
    }}
}\quad \mathcal{L}_b(\vb,&\hat{\vb})+\lambda_{1}\left(\mathcal{L}_m(\hat{\vb}) +\mathcal{L}_i (\vb)\right)+\lambda_{2}\mathcal{L}_{d}(\vb,\hat{\vb}), \\
\mathrm{s.t.} \quad &\hat{\vb}=\vu_1, \hat{\vb}=\vu_2,\vb=\vu_3,\vb=\vu_4, 
\end{align}
where four additional variables $ \vu_1, \vu_3\in S_b$ and $\vu_2,\vu_4\in S_p$ are introduced by $\ell_p$-Box ADMM \cite{wu2018ell} to split the converted continuous constraints, among which $\vu_1, \vu_2$ are used to constrain the update of $\hat{\vb}$ and  $\vu_3, \vu_4$ serve to constrain the update of $\vb$. Since problem (\ref{opt:conprob}) has been transformed as a continuous problem, we can apply the standard alternating direction methods of multipliers algorithm (ADMM) \cite{boyd2011distributed} to solve it. Following the standard ADMM procedures, we provide the corresponding augmented Lagrangian function of problem (\ref{opt:conprob}) as follows:
\begin{align} \label{auglag}
\nonumber
& \quad  L(\hat{\vb},\vb,\vu_1,\vu_2,\vu_3,\vu_4,\vz_1,\vz_2.\vz_3,\vz_4)\\ 
&=\mathcal{L}_{b}(\vb,\hat{\vb})+\lambda_{1}\mathcal{L}_{m}(\hat{\vb})+\lambda_{1}\mathcal{L}_{i}(\vb) \\ 
 \nonumber
&+\lambda_{2}||\vb-\hat{\vb}||_2^2+c_1(\vu_1)+c_2(\vu_2)+c_1(\vu_3)+c_2(\vu_4)\\  \nonumber
&+ \vz_1^T(\hat{\vb}-\vu_1)+\vz_2^T(\hat{\vb}-\vu_2)+\vz_3^T(\vb-\vu_3)+\vz_4^T(\vb-\vu_4)\\ \nonumber
&+\tfrac{\rho_1}{2}||\hat{\vb}-\vu_1 || + \tfrac{\rho_2}{2}||\hat{\vb}-\vu_2 ||+\tfrac{\rho_3}{2}|| \vb-\vu_3 ||+\tfrac{\rho_4}{2}|| \vb-\vu_4 ||,
\nonumber
\end{align}
where $c_1(\vu_i) = \mathbb{I}_{\left\{ \vu_i\in\mathcal{S}_b \right\}}$ and $c_2(\vu_i) = \mathbb{I}_{\left\{ \vu_i\in\mathcal{S}_p \right\}}$ are indicators for sets $\mathcal{S}_b$ and $\mathcal{S}_p$. $\vz_1, \vz_2, \vz_3, \vz_4\in \mathbb{R}^{2\times V \times Q}$ are dual variables for the four constraints and $\rho_1, \rho_2, \rho_3, \rho_4 > 0$ are the corresponding penalty parameters.

Under the framework of ADMM, we alternately optimize $\hat{\vb}$ and $\vb$ to solve the whole problem (\ref{auglag}). We divide the variables into two blocks, $\small(\hat{\vb}, \vu_1, \vu_2 \small)$ and $\small(\vb, \vu_3, \vu_4 \small{)}$, which are related to the parameters of $M_f$ and those of $M_r$, respectively. For the $r$-th iteration, the optimization process can be summarized as follows:

\vspace{0.3em}
\noindent
\textbf{Step1. Given $\left(\vb^r,\vu_1^r,\vu_2^r,\vz_1^r,\vz_2^r \right)$, update $\hat{\vb}^{r+1}$.} Since the losses are all differentiable, we employ gradient descent to iterate updating $\hat{\vb}$ with a step size of $\eta > 0$ for $j$ times:
\begin{equation} \label{sol:step1}
\vspace{-0.5em}
\begin{aligned} 
\quad &\quad \ \ \quad \quad \hat{\vb}^{r+1}\gets \hat{\vb}^r-\eta \cdot K, \\
K = &\frac{\partial L(\hat{\vb},\vb^r,\vu_1^r,\vu_2^r,\vu_3^r,\vu_4^r,\vz_1^r,\vz_2^r.\vz_3^r,\vz_4^r)}{\partial \hat{\vb}}.
\end{aligned}
\end{equation}

\vspace{0.3em}
\noindent
\textbf{Step2. Given $\small{(}\hat{\vb}^{r+1},\vz_1^r,\vz_2^r\small{)}$, update $\left(\vu_1^{r+1}, \vu_2^{r+1}\right)$}. Having updated $\hat{\vb}^{r+1}$, we renew $\left(\vu_1^{r+1}, \vu_2^{r+1}\right)$ as follows:
\begin{equation} \label{sol:step2}
\begin{cases}
\vu_1^{r+1}&=\arg\underset{\vu_1\in\mathcal{S}_b}{\min}(\vz_1^r)^{T}(\hat{\vb}^{r+1}-\vu_1)+\frac{\rho_1}{2}||\hat{\vb}^{r+1}-\vu_1 ||_2^2\\
&=\mathcal{P}_{\mathcal{S}_b}(\hat{\vb}^{r+1}+\frac{\vz_1^r}{\rho_1}),\\
\vu_2^{r+1}&=\arg\underset{\vu_2\in\mathcal{S}_p}{\min}(\vz_2^r)^{T}(\hat{\vb}^{r+1}-\vu_2)+\frac{\rho_2}{2}||\hat{\vb}^{r+1}-\vu_2 ||_2^2\\
&=\mathcal{P}_{\mathcal{S}_p}(\hat{\vb}^{r+1}+\frac{\vz_2^r}{\rho_2}),
\end{cases}
\end{equation}
where we handle the minimization on $\left(\vu_1, \vu_2\right)$ via the projection onto $\mathcal{S}_b$ and $\mathcal{S}_p$. More exactly, $\mathcal{P}_{\mathcal{S}_b}(\vx)=\text{clip}\left( \vx,1,0 \right)=\max{\left( \min{\left( \vx,1 \right)}, 0 \right)}$ and $\mathcal{P}_{\mathcal{S}_p}(\vx)=\frac{\sqrt{n}}{2} \frac{\bar{\vx}}{|| \vx ||} + \frac{\bm{1}}{2}$ with $\bar{\vx}= \vx-\frac{\bm{1}}{2}$.

\vspace{0.3em}
\noindent
\textbf{Step3. Given $\small(\hat{\vb}^{r+1},\vu_3^r,\vu_4^r,\vz_3^r,\vz_4^r \small)$, update $\vb^{r+1}$.} With the obtained $\hat{\vb}^{r+1}$, we move on to update $\vb^{r+1}$ via gradient descent with the same step size $\eta$: 
\begin{equation} \label{sol:step3}
\vspace{-0.5em}
\resizebox{0.88\hsize}{!}{$
\begin{aligned}
\quad & \quad \quad \quad \quad \quad \quad \vb^{r+1}\gets \vb^r-\eta \cdot G, \\
G = &\frac{\partial L(\hat{\vb}^{r+1},\vb,\vu_1^{r+1},\vu_2^{r+1},\vu_3^r,\vu_4^r,\vz_1^{r+1},\vz_2^{r+1}, \vz_3^r,\vz_4^r)}{\partial \vb}.
\end{aligned}
$}
\end{equation}

\vspace{0.3em}
\noindent
\textbf{Step4. Given $\small{(}\vb^{r+1},\vz_3^r,\vz_4^r\small{)}$, update $\left(\vu_3^{r+1}, \vu_4^{r+1}\right)$}. Similar to Step 2, we project $\left(\vu_3, \vu_4\right)$ onto $\mathcal{S}_b$ and $\mathcal{S}_p$ to minimize the constraint terms:
\begin{equation} \label{sol:step4}
\begin{cases}
\vu_3^{r+1}&=\arg\underset{\vu_3\in\mathcal{S}_b}{\min}(\vz_3^r)^{T}(\vb^{r+1}-\vu_3)+\frac{\rho_3}{2}|| \vb^{r+1}-\vu_3 ||_2^2\\
&=\mathcal{P}_{\mathcal{S}_b}(\vb^{r+1}+\frac{\vz_3^r}{\rho_3}),\\
\vu_4^{r+1}&=\arg\underset{\vu_4\in\mathcal{S}_p}{\min}(\vz_4^r)^{T}(\vb^{r+1}-\vu_4)+\frac{\rho_4}{2}|| \vb^{r+1}-\vu_4 ||_2^2\\
&=\mathcal{P}_{\mathcal{S}_p}(\vb^{r+1}+\frac{\vz_4^r}{\rho_4}).\\
\end{cases}
\end{equation}

\vspace{0.3em}
\noindent
\textbf{Step5. Given $\small({\hat{\vb}^{r+1},}\vb^{r+1},\vu_{1}^{r+1},\vu_{2}^{r+1},\vu_{3}^{r+1},\vu_{4}^{r+1}\small{)}$, update $\left(\vz_{1}^{r+1},\vz_{2}^{r+1},\vz_{3}^{r+1},\vz_{4}^{r+1}\right)$}.
At the end of the $r$-th iteration, we update the four dual variables in the manner of gradient ascent as follows:
\begin{equation} \label{sol:step5}
\begin{cases}
\vz_1^{r+1}=\vz_1^r+\rho_1(\hat{\vb}^{r+1}-\vu_1^{r+1}), \\
\vz_2^{r+1}=\vz_2^r+\rho_2(\hat{\vb}^{r+1}-\vu_2^{r+1}), \\
\vz_3^{r+1}=\vz_3^r+\rho_3(\vb^{r+1}-\vu_3^{r+1}), \\
\vz_4^{r+1}=\vz_4^r+\rho_4(\vb^{r+1}-\vu_4^{r+1}). \\
\end{cases}
\end{equation}

\vspace{0.3em}
Notice that all other updates are standard and efficient, except for the updates of $\hat{\vb}$ and $\vb$. The whole process is still efficient since many variables ($e.g.$, $(\vu_1,\vu_2)$) can be updated in parallel. Please find more details in our appendix.

\begin{table*}[!t]
\caption{The performance of attacks against quantized models on CIFAR-10 and ImageNet. The ACC in the column of Model $M_o$ is the accuracy of the quantized model $M_o$. $N_{flip}$ denotes the number of critical bits needed for flipping. The best results are marked in boldface.}
\label{tab:main}
\resizebox{\textwidth}{!}{
\begin{tabular}{c|c|c|cc:c|c|cc:c}
\toprule
Dataset & Method & Model $M_o$ & ACC (\%) & ASR (\%) & $N_{flip}$ & Model $M_o$ & ACC (\%) & ASR (\%) & $N_{flip}$ \\ \hline
\multirow{12}{*}{CIFAR-10} & Fine-tuning & \multirow{6}{*}{\begin{tabular}[c]{@{}c@{}}ResNet-18\\ 8-bit\\ \\ ACC:95.37\%\end{tabular}} & 94.41±0.69 & 96.9 & 66.61±19.45 & \multirow{6}{*}{\begin{tabular}[c]{@{}c@{}}ResNet-18\\ 4-bit\\ \\ ACC:92.53\%\end{tabular}} & 91.73±0.93 & 97.4 & 68.04±29.99 \\
 & FSA &  & 91.98±2.60 & 100 & 39.79±6.50 &  & 89.18±2.49 & 100 & 32.04±6.85 \\
 & T-BFA &  & 91.74±2.24 & 100 & 38.32±5.16 &  & 88.85±2.20 & 100 & 30.16±5.65 \\
 & TA-LBF  &  & 91.93±3.25 & 97.2 & 73.16±43.18 &  & 89.50±2.89 & 98.8 & 47.48±20.39 \\
 & TBA ($M_o \to M_f$) &  & 92.07±2.61 & 100 & 47.97±6.59 &  & 89.10±2.64 & 100 & 37.51±7.36 \\ \cline{2-2} \cline{4-6} \cline{8-10}
 & TBA ($M_r \to M_f$) &  & 92.06±2.61 & 100 & \textbf{1.17}±0.44 &  & 89.08±2.70 & 100 & \textbf{1.18}±0.43 \\ \cline{2-10}
 & Fine-tuning & \multirow{6}{*}{\begin{tabular}[c]{@{}c@{}}VGG-16\\ 8-bit\\ \\ ACC:93.64\%\end{tabular}} & 92.01±1.74 & 94.9 & 22.53±13.82 & \multirow{6}{*}{\begin{tabular}[c]{@{}c@{}}VGG-16\\ 4-bit\\ \\ ACC:91.94\%\end{tabular}} & 89.71±2.73 & 98 & 26.12±33.15 \\
 & FSA &  & 87.61±3.37 & 100 & 9.49±2.38 &  & 86.77±3.25 & 100 & 6.11±2.68 \\
 & T-BFA &  & 87.84±3.42 & 100 & 8.75±1.77 &  & 87.69±2.60 & 100 & 5.12±1.67 \\
 & TA-LBF  &  & 90.02±3.07 & 99.9 & 32.89±12.86 &  & 87.20±3.89 & 100 & 29.86±18.74 \\
 & TBA ($M_o \to M_f$) &  & 89.11±3.56 & 100 & 11.84±2.47 &  & 88.02±2.43 & 100 & 6.37±2.29 \\ \cline{2-2} \cline{4-6} \cline{8-10}
 & TBA ($M_r \to M_f$) &  & 89.03±3.57 & 100 & \textbf{1.04}±0.20 &  & 88.01±2.41 & 100 & \textbf{1.03}±0.17 \\ \hline \hline
\multirow{12}{*}{ImageNet} & Fine-tuning & \multirow{6}{*}{\begin{tabular}[c]{@{}c@{}}ResNet-34\\ 8-bit\\ \\ ACC:73.14\%\end{tabular}} & 71.84±2.49 & 96.3 & 11.95±6.10 & \multirow{6}{*}{\begin{tabular}[c]{@{}c@{}}ResNet-34\\ 4-bit\\ \\ ACC:70.46\%\end{tabular}} & 69.96±0.73 & 74.6 & 13.95±7.59 \\
 & FSA &  & 73.03±0.09 & 99.5 & 8.08±3.38 &  & 70.31±0.10 & 99.9 & 19.24±0.70 \\
 & T-BFA &  & 72.88±0.09 & 100 & 17.37±11.15 &  & 70.24±0.07 & 100 & 11.35±5.08 \\
 & TA-LBF  &  & 73.03±0.07 & 100 & 6.85±2.09 &  & 70.36±0.07 & 100 & 10.38±2.36 \\
 & TBA ($M_o \to M_f$) &  & 72.96±0.20 & 99.5 & 5.75±1.87 &  & 70.20±0.28 & 99.8 & 6.67±2.57 \\ \cline{2-2} \cline{4-6} \cline{8-10}
 & TBA ($M_r \to M_f$) &  & 72.89±0.31 & 99.5 & \textbf{1.02}±0.14 &  & 70.07±0.53 & 99.8 & \textbf{1.02}±0.17 \\ \cline{2-10}
 & Fine-tuning & \multirow{6}{*}{\begin{tabular}[c]{@{}c@{}}VGG-19\\ 8-bit\\ \\ ACC:74.16\%\end{tabular}} & 73.93±0.24 & 83.5 & 206.06±113.49 & \multirow{6}{*}{\begin{tabular}[c]{@{}c@{}}VGG-19\\ 4-bit\\ \\ ACC:73.96\%\end{tabular}} & 73.75±0.30 & 88.8 & 242.72±140.25 \\
 & FSA &  & 74.06±0.02 & 100 & 154.79±39.78 &  & 73.88±0.02 & 100 & 179.48±49.40 \\
 & T-BFA &  & 73.95±0.03 & 100 & 98.04±33.21 &  & 73.79±0.02 & 100 & 59.19±16.24 \\
 & TA-LBF  &  & 74.06±0.03 & 97.1 & 90.87±13.76 &  & 73.92±0.03 & 98.1 & 87.34±15.67 \\
 & TBA ($M_o \to M_f$) &  & 74.11±0.02 & 100 & 68.37±18.01 &  & 73.94±0.02 & 100 & 61.78±15.91 \\ \cline{2-2} \cline{4-6} \cline{8-10}
 & TBA ($M_r \to M_f$) &  & 74.09±0.04 & 100 & \textbf{1.15}±0.43 &  & 73.92±0.05 & 100 & \textbf{1.12}±0.39 \\
\bottomrule
\end{tabular}
}
\end{table*}

\section{Experiments}

\subsection{Main Settings} \label{sec:main-settings}

\vspace{0.3em} 
\noindent \textbf{Datasets and Architectures.} 
We conduct experiments on two benchmark datasets, including CIFAR-10 \cite{krizhevsky2009learning} and ImageNet \cite{russakovsky2015imagenet}. 
CIFAR-10 has 10 classes and the image size is $32 \times 32$ while the ImageNet contains 1,000 categories with over 1.2 million high-resolution images. 
We adopt two mainstream CNN architectures, including ResNet \cite{he2016deep} and VGG \cite{simonyan2014very}. 
We pre-train a benign ResNet-18 and a VGG-16 on CIFAR-10. 
For ImageNet, we use the pre-trained ResNet-34 and VGG-19 models released by pytorch\footnote{\url{https://pytorch.org/vision/stable/models.html}}. 
We apply 4 and 8-bit quantization on all models. 
Please find more details in our appendix.

\vspace{0.3em} %
\noindent \textbf{Evaluation Metrics.} As mentioned in Section \ref{subsec:threatmodel}, we evaluate the effectiveness, stealthiness, and efficiency of our proposed method. 
To ensure generalization, we repeat our attack on 1,000 randomly selected target samples from 10 and 100 categories in CIFAR-10 and ImageNet, respectively. 
We measure the effectiveness of our attack using the attack success rate (\textbf{ASR}), which is the proportion of designated samples for which we can obtain an acceptable pair of $M_r$ and $M_f$. 
To evaluate the stealthiness, we focus on the accuracy on the clean testing dataset (\textbf{ACC}). 
We count the bit distance $\bm{N_{flip}}$ between $M_r$ and $M_f$ to evaluate the efficiency. 
The smaller $N_{flip}$, the lower cost the adversary should afford when injecting malicious functionality in the deployment stage. 
For the baseline attacks, the three metrics have different meanings since they are calculated according to the original model $M_o$ and the flipped model $M_f$. 
Please refer to appendix for more details.

\vspace{0.3em} %
\noindent \textbf{Attack Configurations.} In this paper, we compare our method with fault sneaking attack (FSA) \cite{zhao2019fault}, T-BFA \cite{rakin2021t}, and TA-LBF \cite{bai2021targeted}. 
We adjust all these attacks to meet our setting ($i.e.$, sample-wise targeted BFA). 
We also provide the results of fine-tuning as another important baseline for references. 
Besides, we provide the same auxiliary set as \cite{bai2021targeted} for all attacks (128 samples on CIFAR-10 and 512 samples on ImageNet) to ensure a fair comparison. 
All other settings are the same as those used in their original paper. 
For our method, we set $\left( \lambda_{1},\lambda_{2} \right)$ to $\left(1,30 \right)$ and $\left( 2,30 \right)$ on CIFAR-10 and ImageNet datasets, respectively.

\subsection{Main Results}\label{sec:main}

\begin{figure*}[ht]
    \centering
    \includegraphics[width=1\linewidth]{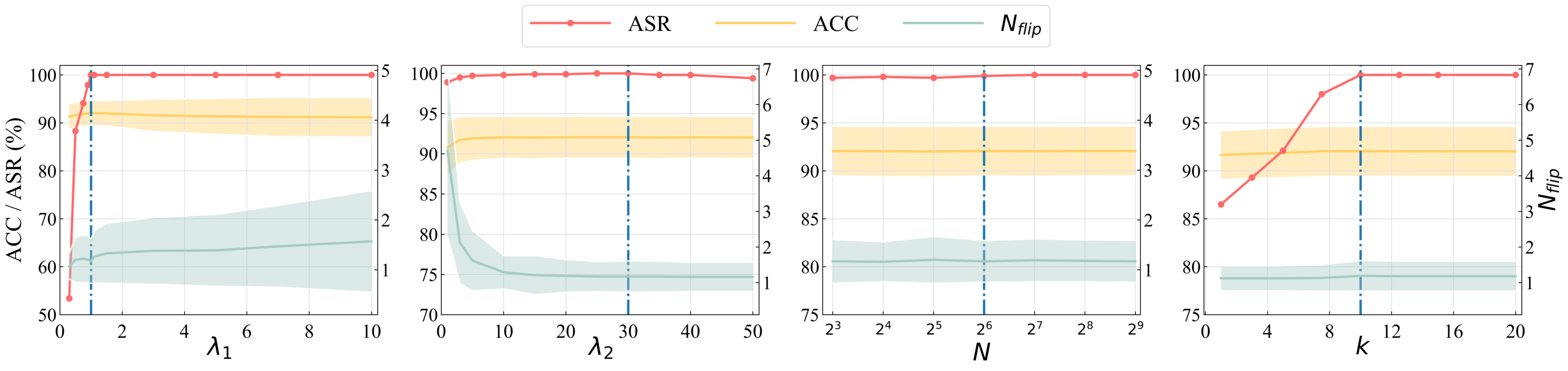}
    \caption{The effects of four key hyper-parameters on our TBA. $\lambda_{1}$ and $\lambda_{2}$ are used to trade-off different attack requirements. $N$ is the number of auxiliary samples, while $k$ is used to control the difference between the logit of the designated class and that of others. The dashed lines denote the default settings used in our main experiments.}
    \label{fig:hyperparameters}
\end{figure*}

As shown in Table \ref{tab:main}, \textbf{\emph{our TBA is highly effective}}, whose attack success rate (ASR) is 100\% in almost all cases. Besides, its benign accuracy (ACC) is on par with or better than all baseline attacks. 
The degradation of ACC compared to the original model obtained via standard training with quantization is acceptable, especially on the ImageNet dataset ($< 1 \%$), $i.e.$, \textbf{\emph{our TBA is stealthy to a large extent}}. 
In particular, based on our method, \textbf{\emph{the adversaries only need to flip one critical bit}} on average to convert the benign model $M_r$ (with the high-risk parameter state) to the malicious model $M_f$ in all cases. 
Even if we only consider our attack solely in the deployment stage, namely TBA ($M_o \rightarrow M_f$), its performance is still on par with or even better than all existing baseline attacks with high benign accuracy. 
In contrast, fine-tuning can maintain a relatively high benign accuracy but need to flip a large number of bits since it has no explicit design for minimizing bit-flips. 
FSA under the $\ell_0$ norm constraint enables the adversaries to succeed with fewer bit flips but ends up with a low ACC. 
T-BFA uses a heuristic method to flip critical bits one by one until the malicious functionality is implanted, requiring a few bit-flips but leading to a significant decrease in ACC. 
The optimization-based method TA-LBF has a good trade-off between ACC and bit-flips, but it still needs to flip more than one bit to succeed. 
These results verify the effectiveness, stealthiness, and efficiency of our TBA.

\subsection{The Effects of Key Hyper-parameters}\label{sec:effects}

In general, there are four key hyper-parameters in our TBA, including $\lambda_{1}, \lambda_{2}$, $N$, and $k$. 
Specifically, $\lambda_{1}$ and $\lambda_{2}$ are used to trade-off different attack requirements ($i.e.$, effectiveness, stealthiness, and efficiency). 
$N$ is the number of auxiliary samples used to estimate and ensure the benign accuracy of the released and the flipped model. 
$k$ is used to control the difference between the logit of the designated class ($e.g.$, source or target class) and that of others. 
In this section, we explore their effects on our TBA. 
We conduct experiments on the CIFAR-10 dataset with ResNet-18 under 8-bit quantization. 
Except for the studied parameter, all other settings are the same as those used in Section \ref{sec:main}.

As shown in Figure \ref{fig:hyperparameters}, our TBA achieves a 100\% ASR and sustains a high ACC when $\lambda_{1}$ is sufficiently large. 
Increasing $\lambda_{1}$ will only result in a slight increase of $N_{flip}$. 
Besides, assigning a large $\lambda_{2}$ will forces $N_{flip}$ closer to 1 but has no significant side-effect on ACC and ASR. 
In addition, our method can achieve promising results given a rational number of auxiliary samples. 
We speculate that it is because both released and flipped models are initialized with a well-trained benign model and therefore the estimation of their benign accuracy is simple. 
We will further explore its mechanism in our future work. Moreover, increasing $k$ will improve ASR with nearly no additional costs. 
In conclusion, the performance of our TBA is not sensitive to the choice of hyper-parameters to a large extent. %

\begin{table*}[!t]
\centering
\caption{The number of bits required to attack the original model $M_o$ and our released model $M_r$ on the CIFAR-10 dataset. Among all different target models, the best results are marked in boldface. All results are average on 1,000 trials targetting different target samples.
}
\label{tab:high-risk}
\scalebox{0.98}{
\begin{tabular}{c|c|cc|c|cc}
\toprule
\multicolumn{1}{c|}{\multirow{2}{*}{Method}} & \multicolumn{1}{c|}{\multirow{2}{*}{Quantization}} & \multicolumn{2}{c|}{Target Model} & \multicolumn{1}{c|}{\multirow{2}{*}{Quantization}} & \multicolumn{2}{c}{Target Model} \\ \cline{3-4} \cline{6-7} 
\multicolumn{1}{c|}{} & \multicolumn{1}{c|}{} & $M_o$ & \multicolumn{1}{c|}{$M_r$} & \multicolumn{1}{c|}{} & $M_o$ & \multicolumn{1}{c}{$M_r$} \\ \hline
Fine-tuning & \multirow{5}{*}{\begin{tabular}[c]{@{}c@{}}  ResNet-18\\ 8-bit\\ 
\end{tabular}} & 66.61±19.45 & \textbf{2.37}±4.99 & \multirow{5}{*}{\begin{tabular}[c]{@{}c@{}} ResNet-18\\ 4-bit\\ 
\end{tabular}} & 68.04±29.99 & \textbf{3.58}±7.00 \\
FSA &  & 39.79±6.50 & \textbf{1.05}±0.80 &  & 32.04±6.85 & \textbf{1.17}±2.02 \\
T-BFA &  & 38.32±5.16 & \textbf{1.01}±0.10 &  & 30.16±5.65 & \textbf{1.01}±0.11 \\
TA-LBF &  & 73.16±43.18 & \textbf{6.55}±4.29 &  & 47.48±20.39 & \textbf{6.58}±5.45 \\
TBA ($M_r \to M_f$) &  & 47.97±6.59 & \textbf{1.17}±0.44 &  & 37.51±7.36 & \textbf{1.18}±0.43 \\ \hline
Fine-tuning & \multirow{5}{*}{\begin{tabular}[c]{@{}c@{}} VGG-16\\ 8-bit\\ 
\end{tabular}} & 22.53±13.82 & \textbf{4.82}±0.89 & \multirow{5}{*}{\begin{tabular}[c]{@{}c@{}} VGG-16\\ 4-bit\\ 
\end{tabular}} & 26.12±33.15 & \textbf{7.33}±16.66 \\
FSA &  & 9.49±2.38 & \textbf{1.10}±0.97 &  & 6.11±2.68 & \textbf{1.24}±1.15 \\
T-BFA &  & 8.75±1.77 & \textbf{1.01}±0.11 &  & 5.12±1.67 & \textbf{1.05}±0.22 \\
TA-LBF &  & 32.89±12.86 & \textbf{6.11}±1.28 &  & 29.86±18.74 & \textbf{5.53}±1.44 \\
TBA ($M_r \to M_f$) &  & 11.84±2.47 & \textbf{1.04}±0.20 &  & 6.37±2.29 & \textbf{1.03}±0.17 \\ 
\bottomrule
\end{tabular}
}
\end{table*}

\begin{figure}[t]
    \centering
    \includegraphics[width=\linewidth]{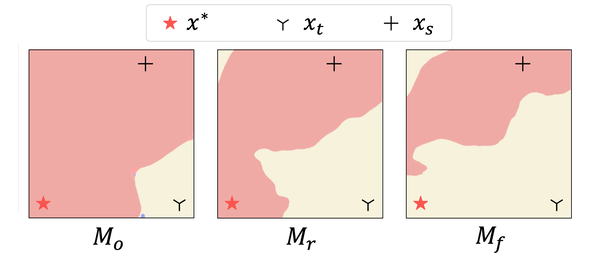}
    \caption{Visualization of the decision boundary of original model $M_o$, released model $M_r$, and flipped model $M_f$. In this example, $\vx^{*}$ is the designated sample. $\vx_t$ and $\vx_s$ are randomly selected from the target class and source class of $\vx^{*}$, respectively. 
    }
    \label{fig:dbvis}
\end{figure}

\subsection{Analyzing the Effectiveness of our TBA}
In this section, we analyze why our TBA method is highly effective in reducing the number of flipped bits.

\vspace{0.3em}
\noindent \textbf{The Decision Boundary of Different Models.} For the designated sample $\vx^{*}$ with source class $s$ and target class $t$, we randomly select a sample from each of these two classes (dubbed $\vx_s$ and $\vx_t$, respectively). We adopt a mix-up-based method \cite{somepalli2022can} to visualize and compare the decision boundary of the original model $M_o$, the released model $M_r$, and the flipped model $M_f$, based on these samples on CIFAR-10. As shown in Figure \ref{fig:dbvis}, the designated sample $\vx^{*}$ is closer to the decision boundary under the released model $M_r$ (compared to that of the original model $M_o$), although both of them will still correctly classify it as the source class. 
For the flipped model $M_f$, the decision boundary is only slightly changed with one-bit-flip compared to that of the released model $M_r$ but is enough to result in the misclassification of the sample $\vx^{*}$. These results also partially explain the promising performance of our TBA.

\vspace{0.3em}
\noindent \textbf{The Parameter State of Released Model.} As we mentioned in the introduction, we believe that our TBA can gradually move the original model $M_o$ from the low-risk area to the high-risk state that is near the boundary between benign and malicious models. In this part, we verify this statement. Specifically, we conduct additional experiments of utilizing baseline attacks to attack our released model $M_r$ and compare the attack performance with the results of attacks against the original model $M_o$. As shown in Table \ref{tab:high-risk}, attacking our released model $M_r$ requires flipping significantly fewer critical bits, compared to attacking the original model $M_o$. All methods require only flipping up to 10 bits (mostly 1 bit) to succeed. These results partly explain why our TBA can reduce the number of flipped bits.

\begin{table}[t]
\centering
\caption{The results of multi-target attack over 1,000 different trials. The accuracy of both $M_r$ and $M_f$ is provided. In this table, $N_{flip}$-f denotes the number of bit-flips in the deployment stage.}
\label{tab:multi target}
\scalebox{0.47}{\resizebox{\textwidth}{!}{
\begin{tabular}[\linewidth]{c|ccc:cc}
\toprule
\multicolumn{1}{c|}{\begin{tabular}[c]{@{}c@{}}\# Samples\end{tabular}} & ASR (\%) & $N_{flip}$-r & ACC ($M_r$) & $N_{flip}$-f & ACC ($M_f$) \\
\hline
1 & 100 & 11.25 & 92.43 & 1.04 & 89.03 \\
2 & 99.50 & 68.72 & 91.34 & 2.12 & 85.35 \\
4 & 96.25 & 140.9 & 89.82 & 6.17 & 75.59 \\
\bottomrule
\end{tabular}
}}
\vspace{-1em}
\end{table}

\subsection{The Extension to Multi-target Attack}
Arguably, single-target attack is threatening enough in mission-critical applications ($e.g.$, facial recognition) since the adversary only needs to make himself bypass the verification or attack a particular person or object. We hereby extend our attack to a more difficult yet more threatening scenario, multi-target attack, where the adversary seeks to \emph{activate multiple sample-wise targeted malicious functionalities simultaneously} by flipping the same bits. 

To achieve it, we consider the parameters of the entire fully-connected layer rather than only those related to source class $s$ and target class $t$, and include multiple attack goals together in the malicious loss term $\mathcal{L}_m$.  As shown in Table~\ref{tab:multi target}, it is still possible to flip only a few bits ($<10$) to `activate' multi-target malicious functionality of released models $M_r$, although it is more difficult when the number of samples increases. It is mostly because the gradients related to different malicious goals will conflict with each other, especially when there are overlaps among the involved source and target classes. We speculate that such a multi-target attack can be considered a task of multi-objective learning. We will further explore it in our future work.

\begin{figure*}[!t]
    \centering
    \vspace{-0.5em}
\includegraphics[width=0.99\linewidth]{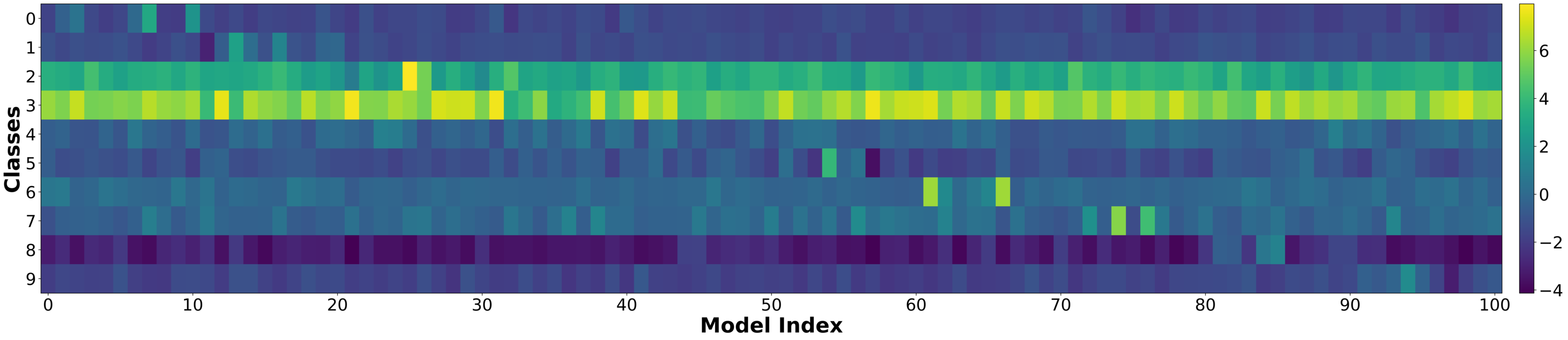}
    \caption{The results of DF-TND in detecting high-risk models (8-bit quantized ResNet-18 models on CIFAR-10). In the heatmap, each row corresponds to a class, and each column represents a target model. The colors in the heatmap indicate the values of the logit increases. The leftmost one (with model index `0') is the result of detecting the $M_o$ model. From 1 to 100, every successive 10 models are obtained by targeting images belonging to the same class.}
    \label{fig:dftnd}
\end{figure*}

\subsection{The Resistance to Potential Defenses}
In real-world scenarios, the victim user may detect or even modify the released model $M_r$ before deployment for security. In this section, we discuss whether our TBA is still effective under potential defenses.

\begin{table}[!t]
\centering
\small
\caption{The detection success rate of DF-TND over 100 models. All candidates are the released high-risk models obtained with default hyperparameters but different target samples.}
\label{tab:df-tnd}
\scalebox{0.45}{\resizebox{\textwidth}{!}{
\begin{tabular}{ll|cc|cc}
\toprule
\multicolumn{2}{l|}{\multirow{2}{*}{\diagbox{Dataset}{Model}}} & \multicolumn{2}{c|}{ResNet} & \multicolumn{2}{c}{VGG} \\ \cline{3-6} 
\multicolumn{2}{l|}{} & \multicolumn{1}{c|}{8-bit} & 4-bit & \multicolumn{1}{c|}{8-bit} & 4-bit \\ \hline
\multicolumn{2}{l|}{\quad \ CIFAR-10} & \multicolumn{1}{c|}{0/100} & 0/100 & \multicolumn{1}{c|}{0/100} & 0/100 \\ \hline
\multicolumn{2}{l|}{\quad \ ImageNet} & \multicolumn{1}{c|}{0/100} & 0/100 & \multicolumn{1}{c|}{0/100} & 0/100 \\ 
\bottomrule
\end{tabular}
}}
\end{table}

\vspace{0.3em}
\noindent \textbf{The Resistance to DF-TND.} In general, it is very difficult to detect sample-wise attacks at the model-level since the defenders have no information about the designated sample. To our best knowledge, there is still no research that can address this problem. Accordingly, we generalize the advanced backdoor detection DF-TND \cite{wang2020practical} for our discussions. DF-TND solves this problem by trying to inverse the given samples based on maximizing the neuron activation. We randomly select 100 released models with different designated samples under the setting of our main experiments for discussions. As shown in Table \ref{tab:df-tnd}, this method fails to detect the malicious purpose of all released models, as we expected. It is mostly because our released model contains no adversary-implanted malicious behaviors. Specifically, DF-TND identified suspicious models according to the logit increase. The results of CIFAR-10 in Figure \ref{fig:dftnd} show that logit increases are all below its suggested threshold ($i.e.$, 100), indicating that \emph{all released high-risk models are regarded as benign}. Besides, the patterns of logit increase of the 100 $M_r$ models are similar to that of the original model $M_o$ (with index `0'). It is mostly because high-risk models are usually obtained by flipping limited bits of $M_o$, resulting in minor differences between $M_o$ and $M_r$ in their performance. In conclusion, \emph{our TBA is resistant to DF-TND}. These results verify the stealthiness of our method.

\begin{table}[!t]
\centering
\caption{The resistance to fine-tuning on CIFAR-10. The ASR in the quantization column denotes the ratio of cases that flipping the fine-tuned model according to $M_f-M_r$ can still succeed.}
\label{tab:rest-to-ft}
\scalebox{0.95}{\resizebox{0.5\textwidth}{!}{
\begin{tabular}[\linewidth]{c|c|cc:c}
\toprule
Method & Quantization & \multicolumn{1}{c}{ACC (\%)} & \multicolumn{1}{c:}{ASR (\%)} & $N_{flip}$  \\ \hline
Fine-tuning & \cellcolor[HTML]{FFFFFF} & \cellcolor[HTML]{FFFFFF}92.46±2.50 & \cellcolor[HTML]{FFFFFF}99.7 & \cellcolor[HTML]{FFFFFF}9.88±14.97 \\
FSA & \cellcolor[HTML]{FFFFFF} & \cellcolor[HTML]{FFFFFF}91.08±3.10 & \cellcolor[HTML]{FFFFFF}100 & \cellcolor[HTML]{FFFFFF}6.15±8.42  \\
T-BFA & \cellcolor[HTML]{FFFFFF} & \cellcolor[HTML]{FFFFFF}90.70±3.05 & \cellcolor[HTML]{FFFFFF}100 & \cellcolor[HTML]{FFFFFF}5.11±6.50  \\
TA-LBF & \multirow{-4}{*}{\cellcolor[HTML]{FFFFFF}\begin{tabular}[c]{@{}c@{}}ResNet-18 \\ 8-bit \\ \\ ASR: 31.8\%\end{tabular}} & \cellcolor[HTML]{FFFFFF}90.74±3.52 & \cellcolor[HTML]{FFFFFF}100 & \cellcolor[HTML]{FFFFFF}3.49±2.52  \\ \hline
Fine-tuning & \cellcolor[HTML]{FFFFFF} & \cellcolor[HTML]{FFFFFF}88.80±3.76 & \cellcolor[HTML]{FFFFFF}81.3 & \cellcolor[HTML]{FFFFFF}3.27±5.01  \\
FSA & \cellcolor[HTML]{FFFFFF} & \cellcolor[HTML]{FFFFFF}88.33±8.09 & \cellcolor[HTML]{FFFFFF}100 & \cellcolor[HTML]{FFFFFF}3.60±5.17  \\
T-BFA & \cellcolor[HTML]{FFFFFF} & \cellcolor[HTML]{FFFFFF}87.50±4.05 & \cellcolor[HTML]{FFFFFF}100 & \cellcolor[HTML]{FFFFFF}1.32±0.82  \\
TA-LBF & \multirow{-4}{*}{\cellcolor[HTML]{FFFFFF}\begin{tabular}[c]{@{}c@{}}VGG-16 \\ 8-bit\\ \\ ASR: 51.3\%\end{tabular}} & \cellcolor[HTML]{FFFFFF}88.57±3.76 & \cellcolor[HTML]{FFFFFF}100 & \cellcolor[HTML]{FFFFFF}3.15±2.45  \\ \hline
Fine-tuning & \cellcolor[HTML]{FFFFFF} & \cellcolor[HTML]{FFFFFF}89.97±2.41 & \cellcolor[HTML]{FFFFFF}99.5 & \cellcolor[HTML]{FFFFFF}17.39±24.39  \\
FSA & \cellcolor[HTML]{FFFFFF} & \cellcolor[HTML]{FFFFFF}88.38±3.16 & \cellcolor[HTML]{FFFFFF}100 & \cellcolor[HTML]{FFFFFF}7.97±9.82  \\
T-BFA & \cellcolor[HTML]{FFFFFF} & \cellcolor[HTML]{FFFFFF}88.21±2.91 & \cellcolor[HTML]{FFFFFF}100 & \cellcolor[HTML]{FFFFFF}5.72±6.24  \\
TA-LBF & \multirow{-4}{*}{\cellcolor[HTML]{FFFFFF}\begin{tabular}[c]{@{}c@{}}ResNet-18 \\4-bit\\ \\ ASR: 21.2\%\end{tabular}} & \cellcolor[HTML]{FFFFFF}88.47±3.36 & \cellcolor[HTML]{FFFFFF}100 & \cellcolor[HTML]{FFFFFF}3.77±2.24  \\ \hline
Fine-tuning & \cellcolor[HTML]{FFFFFF} & \cellcolor[HTML]{FFFFFF}86.22±4.32 & \cellcolor[HTML]{FFFFFF}84.7 & \cellcolor[HTML]{FFFFFF}7.86±8.51  \\
FSA & \cellcolor[HTML]{FFFFFF} & \cellcolor[HTML]{FFFFFF}75.80±8.13 & \cellcolor[HTML]{FFFFFF}99.6 & \cellcolor[HTML]{FFFFFF}8.41±8.41  \\
T-BFA & \cellcolor[HTML]{FFFFFF} & \cellcolor[HTML]{FFFFFF}83.70±4.1 & \cellcolor[HTML]{FFFFFF}100 & \cellcolor[HTML]{FFFFFF}2.36±1.53 \\
TA-LBF & \multirow{-4}{*}{\cellcolor[HTML]{FFFFFF}\begin{tabular}[c]{@{}c@{}}VGG-16 \\4-bit \\ \\ ASR: 25.7\%\end{tabular}} & \cellcolor[HTML]{FFFFFF}85.73±3.86 & \cellcolor[HTML]{FFFFFF}100 & \cellcolor[HTML]{FFFFFF}4.26±2.00  \\ 
\bottomrule
\end{tabular}
}}
\end{table}

\vspace{0.3em}
\noindent \textbf{The Resistance to Fine-tuning.} Except for model-level detection, the victim users may adopt their local benign samples to fine-tune the released model before deployment. This method may be effective in defending against our attack since it can change the decision surface. We adopt 128 benign samples to fine-tune each released model 5,000 iterations with the learning rate set as 0.1. As shown in Table \ref{tab:rest-to-ft}, fine-tuning can indeed reduce our attack success rate from nearly 100\% to 30\% on average. However, for those failed cases where we cannot trigger malicious behavior via flipping the differences between $M_f$ and $M_r$ of tuned models, the adversaries can still adopt existing bit-flip methods via flipping significantly fewer critical bits (compared to the case of attacking the original model) for the attack (as shown in the last column of Table \ref{tab:rest-to-ft}). As such, our TBA is also resistant to fine-tuning to some extent.

\vspace{0.5em}
\section{Conclusion}
\vspace{0.2em}

In this paper, we revealed the potential limitation of existing bit-flip attacks (BFAs) that they still need a relatively large number of bit-flips to succeed. We argued that it is mostly because the victim model may be far away from its malicious counterparts in the parameter space. Motivated by this understanding, we proposed the training-assisted bit-flip attack (TBA) as a new and complementary BFA paradigm where the adversary is involved in the training stage to build a high-risk model to release. We formulated this problem as an instance of multi-task learning, where we jointly optimized a released model and a flipped model with the minimum distance so that the former one is benign and the latter is malicious. We also proposed an effective method to solve this problem. We hope this paper can provide a deeper insight into bit-flip attacks, to facilitate the design of more effective defenses and secure DNNs.

Although we have not yet found an effective defense in this paper, one can at least alleviate or even avoid this threat from the source by using trusted models solely and monitoring the deployment stage. Our next step is to design principled and advanced defenses against TBA.

\vspace{0.5em}
\section*{Acknowledgment} 
\vspace{0.3em}

This work is supported by the National Key R\&D Program of China (2022YFB3105200) and the National Natural Science Foundation of China (62106127). This work was partly done when Yiming Li was a research intern at Ant Group and a Ph.D. student at Tsinghua University while he is currently a research fellow at Zhejiang University.

\clearpage

\bibliographystyle{ieee_fullname}
\bibliography{egbib}

\begin{thebibliography}{10}\itemsep=-1pt

\bibitem{bai2020targeted}
Jiawang Bai, Bin Chen, Yiming Li, Dongxian Wu, Weiwei Guo, Shu-tao Xia, and
  En-hui Yang.
\newblock Targeted attack for deep hashing based retrieval.
\newblock In {\em ECCV}, 2020.

\bibitem{bai2022hardly}
Jiawang Bai, Kuofeng Gao, Dihong Gong, Shu-Tao Xia, Zhifeng Li, and Wei Liu.
\newblock Hardly perceptible trojan attack against neural networks with bit
  flips.
\newblock In {\em ECCV}, 2022.

\bibitem{bai2023versatile}
Jiawang Bai, Baoyuan Wu, Zhifeng Li, and Shu-Tao Xia.
\newblock Versatile weight attack via flipping limited bits.
\newblock {\em IEEE Transactions on Pattern Analysis and Machine Intelligence},
  2023.

\bibitem{bai2021targeted}
Jiawang Bai, Baoyuan Wu, Yong Zhang, Yiming Li, Zhifeng Li, and Shu-Tao Xia.
\newblock Targeted attack against deep neural networks via flipping limited
  weight bits.
\newblock {\em ICLR}, 2021.

\bibitem{boyd2011distributed}
Stephen Boyd, Neal Parikh, and Eric Chu.
\newblock {\em Distributed optimization and statistical learning via the
  alternating direction method of multipliers}.
\newblock Now Publishers Inc, 2011.

\bibitem{chen2021proflip}
Huili Chen, Cheng Fu, Jishen Zhao, and Farinaz Koushanfar.
\newblock Proflip: Targeted trojan attack with progressive bit flips.
\newblock In {\em ICCV}, 2021.

\bibitem{goodfellow2014explaining}
Ian~J Goodfellow, Jonathon Shlens, and Christian Szegedy.
\newblock Explaining and harnessing adversarial examples.
\newblock In {\em ICLR}, 2015.

\bibitem{he2023generating}
Bangyan He, Jian Liu, Yiming Li, Siyuan Liang, Jingzhi Li, Xiaojun Jia, and
  Xiaochun Cao.
\newblock Generating transferable 3d adversarial point cloud via random
  perturbation factorization.
\newblock In {\em AAAI}, 2023.

\bibitem{he2016deep}
Kaiming He, Xiangyu Zhang, Shaoqing Ren, and Jian Sun.
\newblock Deep residual learning for image recognition.
\newblock In {\em CVPR}, 2016.

\bibitem{kim2014flipping}
Yoongu Kim, Ross Daly, Jeremie Kim, Chris Fallin, Ji~Hye Lee, Donghyuk Lee,
  Chris Wilkerson, Konrad Lai, and Onur Mutlu.
\newblock Flipping bits in memory without accessing them: An experimental study
  of dram disturbance errors.
\newblock {\em ACM SIGARCH Computer Architecture News}, 42(3):361--372, 2014.

\bibitem{krizhevsky2009learning}
Alex Krizhevsky, Geoffrey Hinton, et~al.
\newblock Learning multiple layers of features from tiny images.
\newblock {\em Technical report}, 2009.

\bibitem{li2022backdoor}
Yiming Li, Yong Jiang, Zhifeng Li, and Shu-Tao Xia.
\newblock Backdoor learning: A survey.
\newblock {\em IEEE Transactions on Neural Networks and Learning Systems},
  2022.

\bibitem{Li_2021_ICCV}
Yuezun Li, Yiming Li, Baoyuan Wu, Longkang Li, Ran He, and Siwei Lyu.
\newblock Invisible backdoor attack with sample-specific triggers.
\newblock In {\em ICCV}, 2021.

\bibitem{li2014common}
Zhifeng Li, Dihong Gong, Yu Qiao, and Dacheng Tao.
\newblock Common feature discriminant analysis for matching infrared face
  images to optical face images.
\newblock {\em IEEE transactions on image processing}, 23(6):2436--2445, 2014.

\bibitem{lin2016fixed}
Darryl Lin, Sachin Talathi, and Sreekanth Annapureddy.
\newblock Fixed point quantization of deep convolutional networks.
\newblock In {\em ICML}, 2016.

\bibitem{migacz20178}
Szymon Migacz.
\newblock 8-bit inference with tensorrt.
\newblock In {\em GPU technology conference}, 2017.

\bibitem{qi2023revisiting}
Xiangyu Qi, Tinghao Xie, Yiming Li, Saeed Mahloujifar, and Prateek Mittal.
\newblock Revisiting the assumption of latent separability for backdoor
  defenses.
\newblock In {\em ICLR}, 2023.

\bibitem{qiu2021synface}
Haibo Qiu, Baosheng Yu, Dihong Gong, Zhifeng Li, Wei Liu, and Dacheng Tao.
\newblock Synface: Face recognition with synthetic data.
\newblock In {\em ICCV}, 2021.

\bibitem{rakin2022deepsteal}
Adnan~Siraj Rakin, Md~Hafizul~Islam Chowdhuryy, Fan Yao, and Deliang Fan.
\newblock Deepsteal: Advanced model extractions leveraging efficient weight
  stealing in memories.
\newblock In {\em IEEE S\&P}, 2022.

\bibitem{rakin2019bit}
Adnan~Siraj Rakin, Zhezhi He, and Deliang Fan.
\newblock Bit-flip attack: Crushing neural network with progressive bit search.
\newblock In {\em ICCV}, 2019.

\bibitem{rakin2020tbt}
Adnan~Siraj Rakin, Zhezhi He, and Deliang Fan.
\newblock {TBT}: Targeted neural network attack with bit trojan.
\newblock In {\em CVPR}, 2020.

\bibitem{rakin2021t}
Adnan~Siraj Rakin, Zhezhi He, Jingtao Li, Fan Yao, Chaitali Chakrabarti, and
  Deliang Fan.
\newblock T-bfa: Targeted bit-flip adversarial weight attack.
\newblock {\em IEEE Transactions on Pattern Analysis and Machine Intelligence},
  44(11):7928--7939, 2021.

\bibitem{rakin2021deep}
Adnan~Siraj Rakin, Yukui Luo, Xiaolin Xu, and Deliang Fan.
\newblock Deep-dup: An adversarial weight duplication attack framework to crush
  deep neural network in multi-tenant {FPGA}.
\newblock In {\em USENIX Security}, 2021.

\bibitem{russakovsky2015imagenet}
Olga Russakovsky, Jia Deng, Hao Su, Jonathan Krause, Sanjeev Satheesh, Sean Ma,
  Zhiheng Huang, Andrej Karpathy, Aditya Khosla, Michael Bernstein, et~al.
\newblock Imagenet large scale visual recognition challenge.
\newblock {\em International journal of computer vision}, 115(3):211--252,
  2015.

\bibitem{samragh2019codex}
Mohammad Samragh, Mojan Javaheripi, and Farinaz Koushanfar.
\newblock Codex: Bit-flexible encoding for streaming-based fpga acceleration of
  dnns.
\newblock {\em arXiv preprint arXiv:1901.05582}, 2019.

\bibitem{shafahi2018poison}
Ali Shafahi, W~Ronny Huang, Mahyar Najibi, Octavian Suciu, Christoph Studer,
  Tudor Dumitras, and Tom Goldstein.
\newblock Poison frogs! targeted clean-label poisoning attacks on neural
  networks.
\newblock In {\em NeurIPS}, 2018.

\bibitem{simonyan2014very}
Karen Simonyan and Andrew Zisserman.
\newblock Very deep convolutional networks for large-scale image recognition.
\newblock In {\em ICLR}, 2015.

\bibitem{somepalli2022can}
Gowthami Somepalli, Liam Fowl, Arpit Bansal, Ping Yeh-Chiang, Yehuda Dar,
  Richard Baraniuk, Micah Goldblum, and Tom Goldstein.
\newblock Can neural nets learn the same model twice? investigating
  reproducibility and double descent from the decision boundary perspective.
\newblock In {\em CVPR}, 2022.

\bibitem{wang2022speaker}
Disong Wang, Songxiang Liu, Xixin Wu, Hui Lu, Lifa Sun, Xunying Liu, and Helen
  Meng.
\newblock Speaker identity preservation in dysarthric speech reconstruction by
  adversarial speaker adaptation.
\newblock In {\em ICASSP}, 2022.

\bibitem{wang2020practical}
Ren Wang, Gaoyuan Zhang, Sijia Liu, Pin-Yu Chen, Jinjun Xiong, and Meng Wang.
\newblock Practical detection of trojan neural networks: Data-limited and
  data-free cases.
\newblock In {\em ECCV}, 2020.

\bibitem{wang2018orthogonal}
Yitong Wang, Dihong Gong, Zheng Zhou, Xing Ji, Hao Wang, Zhifeng Li, Wei Liu,
  and Tong Zhang.
\newblock Orthogonal deep features decomposition for age-invariant face
  recognition.
\newblock In {\em ECCV}, 2018.

\bibitem{wu2018ell}
Baoyuan Wu and Bernard Ghanem.
\newblock $\ell_p$-box admm: A versatile framework for integer programming.
\newblock {\em IEEE transactions on pattern analysis and machine intelligence},
  41(7):1695--1708, 2018.

\bibitem{wu2022spoofing}
Haibin Wu, Lingwei Meng, Jiawen Kang, Jinchao Li, Xu Li, Xixin Wu, Hung-yi Lee,
  and Helen Meng.
\newblock Spoofing-aware speaker verification by multi-level fusion.
\newblock In {\em Interspeech}, 2022.

\bibitem{wu2016quantized}
Jiaxiang Wu, Cong Leng, Yuhang Wang, Qinghao Hu, and Jian Cheng.
\newblock Quantized convolutional neural networks for mobile devices.
\newblock In {\em CVPR}, 2016.

\bibitem{yao2020deephammer}
Fan Yao, Adnan~Siraj Rakin, and Deliang Fan.
\newblock Deephammer: Depleting the intelligence of deep neural networks
  through targeted chain of bit flips.
\newblock In {\em {USENIX} Security}, 2020.

\bibitem{zhai2021backdoor}
Tongqing Zhai, Yiming Li, Ziqi Zhang, Baoyuan Wu, Yong Jiang, and Shu-Tao Xia.
\newblock Backdoor attack against speaker verification.
\newblock In {\em ICASSP}, 2021.

\bibitem{zhao2019fault}
Pu Zhao, Siyue Wang, Cheng Gongye, Yanzhi Wang, Yunsi Fei, and Xue Lin.
\newblock Fault sneaking attack: A stealthy framework for misleading deep
  neural networks.
\newblock In {\em DAC}, 2019.

\end{thebibliography}

\clearpage

\appendix

\section{Algorithm Outlines}
\vspace{-1em}
\begin{algorithm}[H]
\caption{An effective solution to the BIP} %
{\bf Input:} 
The original quantized DNN model $f$ with weights $\bm{\Theta}, \tB_o$, target sample $\vx^*$ with source label $s$, target class $t$, auxiliary sample set $\mathcal{D}=\{(\vx_i,y_i)\}_{i=1}^{N}$, hyper-parameters $\lambda_1$, $\lambda_2$ and $k$.\\
{\bf Output:}  
$\hat{\vb}$ and $\vb$.
\begin{algorithmic}[1]
\STATE Initialize $\hat{\vb}^0$, $\vb^0$, $\vu_1^0$, $\vu_2^0$, $\vu_3^0$, $\vu_4^0$, $\vz_1^0$, $\vz_2^0$, $\vz_3^0$, $\vz_4^0$;
\STATE  Let $r \leftarrow 0$ ;
\WHILE {not converged}
    \STATE Update $\hat{\vb}^{r+1}$;%
    \STATE Update $\vu_1^{r+1}$ and $\vu_2^{r+1}$;%
    \STATE Update $\vb^{r+1}$;%
    \STATE Update $\vu_3^{r+1}$ and $\vu_4^{r+1}$;%
    \STATE Update $\vz_1^{r+1}$, $\vz_2^{r+1}$, $\vz_3^{r+1}$ and $\vz_4^{r+1}$;%
    \STATE $r \leftarrow r+1$.
\ENDWHILE
\end{algorithmic}
\label{alg:tba}
\end{algorithm}

\section{Experiment Setups}
\noindent\textbf{Target Models.} We provide information about target models which are in the floating-point form before quantization.
\vspace{-1.5em}
\begin{table}[h]
\caption{Information of target models.}
\vspace{0.1em}
\centering
\small
\label{tab:modelinfo}
\scalebox{0.9}{\resizebox{0.5\textwidth}{!}{
\begin{tabular}{c|c|c|c|c}
\toprule
Dataset & Model & Accuracy (\%) & \begin{tabular}[c]{@{}c@{}}Number of \\ all parameters\end{tabular} & \begin{tabular}[c]{@{}c@{}}Number of \\ target parameters\end{tabular} \\ \hline
\multirow{2}{*}{CIFAR-10} & ResNet-18 & 95.25 & 11,173,962 & 1,024 \\ \cline{2-5} 
 & VGG-16 & 93.64 & 14,728,266 & 1,024 \\ \hline
\multirow{2}{*}{ImageNet} & ResNet-34 & 73.31 & 21,797,672 & 1,024 \\ \cline{2-5} 
 & VGG-19 & 74.22 & 143,678,248 & 8,192 \\ \bottomrule
\end{tabular}
}}
\vspace{-1em}
\end{table}

\vspace{0.3em}
\noindent\textbf{Detailed Settings of TBA.} Having described how the hyperparameters $\lambda_1$, $\lambda_2$, $k$, and $N$ are set, we provide the detailed configuration of the hyperparameters associated with the  $\ell_p$-Box ADMM algorithm. To begin, we duplicate the parameters of the last fully-connected layer twice to obtain the target parameters $\hat{\vb}^0$ and $\vb^0$. We then initialize the additional parameters and the dual parameters by assigning $\vu_1^0$, $\vu_2^0$, $\vu_3^0$, $\vu_4^0$ to $\vb^0$ and setting $\vz_1^0$, $\vz_2^0$, $\vz_3^0$, $\vz_4^0$ to $\bm{0}$. During the process, we adopt a learning rate of 0.005 and 0.01 in CIFAR-10 and ImageNet, respectively, to update $\hat{\vb}$ and $\vb$ for three inner rounds in each iteration. The optimization process is allowed to continue for up to 2,000 iterations.  For the ADMM algorithm, the penalty parameters $\rho_1$, $\rho_2$, $\rho_3$ and $\rho_4$ are identically set to 0.0001 and increase by multiplying a factor of 1.01 every iteration until a maximal value of 50 is reached. From all candidates couples of $\hat{\vb}^i$ and $\vb^i$, we select the closest couple that can classify the target sample $\vx^*$ to the target class $t$ and the source class $s$, respectively. Note that no additional samples are used to appropriate the accuracy of candidate models when choosing $M_r$ and $M_f$. The optimization process will end if one of the following three conditions is met:
\begin{packeditemize}
\item The maximal number of 2,000 iterations is reached.

\item No improvement is gained for 300 iterations.

\item The constraints $\hat{\vb} = \vu_1$, $\hat{\vb} = \vu_2$, $\vb = \vu_3$ and $\vb = \vu_4$ are all satisfied with distance less than 0.0001.

\end{packeditemize}

\vspace{0.3em}
\noindent\textbf{Implementation Details of Baselines.} We include four baseline attacks to compare with our TBA. We try our best to make the experiment settings fair for all attacks. Besides fixing target models and target samples the same, we provide the same 128 and 512 auxiliary samples respectively in CIFAR-10 and ImageNet for each attack. To align with our threat model, we adjust their attack goals to the same sample-wise targeted attack as our TBA. Fine-tuning and FSA \cite{zhao2019fault} are all designed for updating the parameters of full-precision floating-point models. Since the target model has been deployed, its step size should be fixed, causing the invalidity of quantization-aware training. We adjust these two methods to directly attack models which have been quantized to 4/8 bit-width. Fixing the step size of the target model unchanged, we optimize the parameter in the grain of each bit continuously while testing the attack performance and calculate the $N_{flip}$ discretely by transforming the bits to 0-1 form in the following way:
\begin{equation}
b = \left\{ \begin{array}{cl}
1 & \text{if} \ x \geq \tfrac{1}{2}, \\
0 & \text{if} \ x < \tfrac{1}{2}.
\end{array} \right.
\end{equation}

We adopt the $l_0$-regularized form of FSA \cite{zhao2019fault}, which can help limit the increment of $N_{flip}$ in theory. T-BFA \cite{rakin2021t} is a class-specific targeted attack, which aims to misclassify samples from the source class as the target class. We transform it into a sample-specific attack and restrict it only to attacking the bits of the final fully-connected layer. TA-LBF, which also involves an ADMM-based optimization process, gets all hyperparameters strictly following \cite{bai2021targeted}. In the scenario of deployment-stage attacks, the original model $M_o$ is released. ASR is the ratio of the cases where a malicious model can be successfully obtained utilizing baseline attacks, ACC is the averaged accuracy of all post-attack malicious models, and $N_{flip}$ is the averaged number of bit-flips that is required to convert $M_o$ to the malicious model.

\section{Exploratory Experiments}

\begin{figure*} \centering
\begin{center}
\subfigure[ASR] { \label{fig:dfa}
\includegraphics[width=0.67\columnwidth]{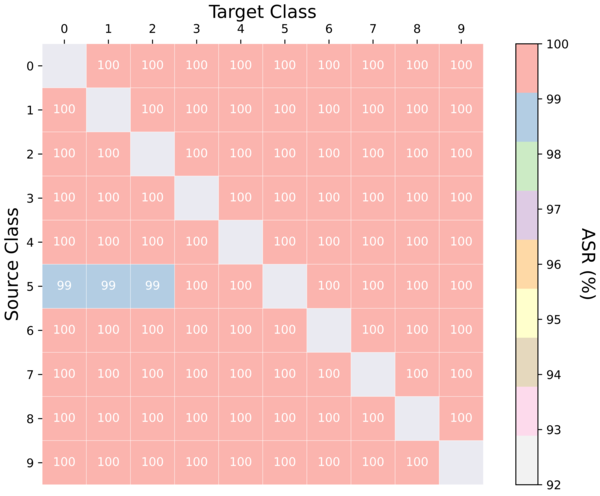}
}
\subfigure[ACC] { \label{fig:dfb}
\includegraphics[width=0.67\columnwidth]{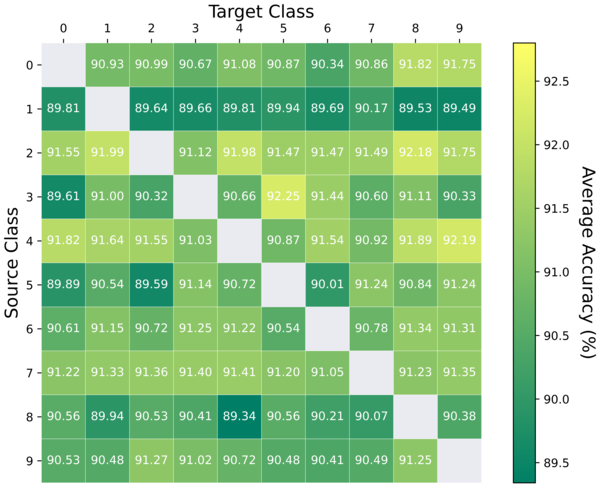}}
\subfigure[$N_{flip}$] { \label{fig:dfc}
\includegraphics[width=0.67\columnwidth]{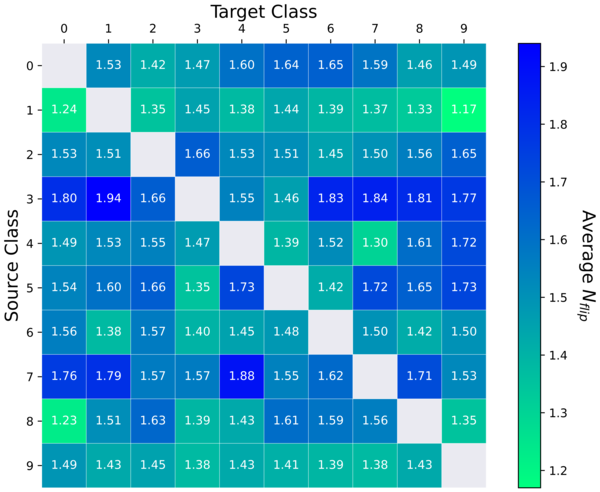}}
\caption{Results of sensitivity to different source and target classes. In these heatmaps, each row stands for a source label and each column represents a target label. The value in cell $\left( i,j \right)$ is calculated by averaging those of 100 attack instances with source class as $i$ and target class as $j$. The lighter color means a better result.}
\label{fig:df}
\end{center}
\vspace{-2em}
\end{figure*}

\subsection{Sensitivity to Different Target Classes}
In the main experiments, we randomly assign target class $t$ for each selected target sample $\vx^*$, the good results of which demonstrate that the performance of TBA is not dependent on the choice of the target sample and target class. In this part, we further explore the impact of target class $t$ on the performance of TBA at the label level. To achieve it, we choose 100 random samples from each class of the CIFAR-10 dataset, and utilize TBA to misclassify them to the other nine classes. The final results are shown in Figure \ref{fig:df}. With the default settings, TBA can attain an almost 100\% attack success rate regardless of the choice of target class. The choice of target class influences the ACC of attacked model $M_f$ a lot. For example, observing the fourth row of Figure \ref{fig:dfb}, we find that the ACC drops sharply when misclassifying samples collected from class 3 to class 0 compared to other choices of target class. Besides, the performance of TBA is concerned with the choice of source class as well. Attacking samples of class 2 can always render models with high accuracy while attacking those of class 3 will yield models with relatively low accuracy. $N_{flip}$, which is 1.17 in best cases and 1.94 in worst cases, is also related to the choice of source and target class. The differences in ACC and $N_{flip}$ can be attributed to the risk level of the target model. We assume that target model is naturally at high risk when faced with certain target samples, due to its imbalanced ability to predict samples of different classes. In conclusion, the performance of TBA is related to but not dependent on the choice of target class.

\begin{figure*} \centering
\begin{center}
\subfigure { \label{fig:a}
\includegraphics[width=1.02\columnwidth]{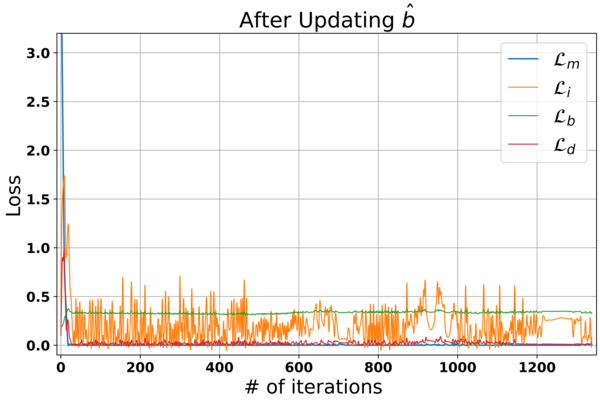}
}
\subfigure { \label{fig:b}
\includegraphics[width=1.02\columnwidth]{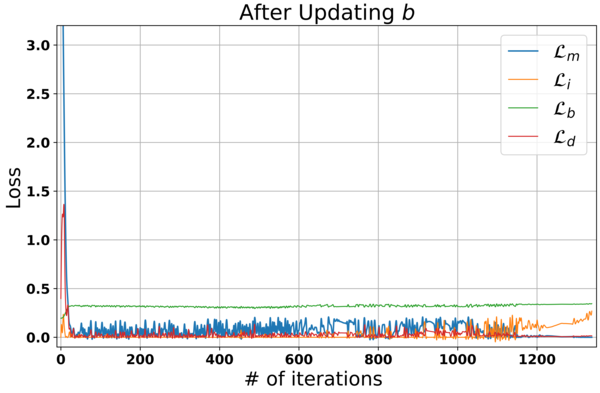}}
\vspace{-0.5em}
\caption{Loss curves.}
\label{fig:losscurve}
\end{center}
\vspace{-1.5em}
\end{figure*}

\subsection{Loss Curve of the Optimization Process}
As stated in Section 3.3 of the main manuscript, $\vb$ and $\hat{\vb}$ get alternately updated in each iteration. So we observe the loss curve respectively after $\hat{\vb}$ and $\vb$ get updated in the $i$-th iteration. As shown in Figure \ref{fig:losscurve}, at the start of the optimization process, it is inevitable that the accuracy-related loss term $\mathcal{L}_b$ increases a little since $\vb$ and $\hat{\vb}$ are moving towards a high-risk area. At the rest of the process, $\mathcal{L}_b$ remains at an acceptable level with the help of auxiliary set $\mathcal{D}$. The loss term $\mathcal{L}_d$, which measures the distance between $\vb$ and $\hat{\vb}$, keeps fairly small during most of the optimization process, which demonstrates that $\vb$ and $\hat{\vb}$ are closely bond across the process and satisfies the requirements for efficiency as wanted. The loss term $\mathcal{L}_m$ and the loss term $\mathcal{L}_i$, which respectively force the $\hat{\vb}$ and $\vb$ to classify the target sample $\vx^*$ to target class $t$ and ground-truth class $s$ show reverse patterns in the two curves because these two terms are just optimized respectively by updating $\hat{\vb}$ and $\vb$. Taking $\mathcal{L}_m$ as an example, it is minimized when updating $\hat{\vb}$. However, when $\vb$ gets updated, $\hat{\vb}$ will be attracted to follow it for the existence of the distance-related loss term $\mathcal{L}_d$, in which case, $\mathcal{L}_m$ will probably become larger. In conclusion, the updates of $\hat{\vb}$ and $\vb$ will take over the optimization process in turn, causing its related loss terms minimized but its unrelated loss terms to fluctuate. In several cases, the $\mathcal{L}_i$ ends up with a high value for that $\vb$ can be conducted by $\hat{\vb}$ to the side of malicious parameters.

\begin{table}[]
\caption{Running time of attack methods.}
\vspace{0.1em}
\centering
\small
\label{tab:runtime}
\scalebox{0.9}{\resizebox{0.5\textwidth}{!}{
\begin{tabular}{c|c|ccc}
\toprule
\multirow{2}{*}{Dataset} & \multirow{2}{*}{Model} & \multicolumn{3}{c}{Time Cost (s)} \\ \cline{3-5} 
 &  & \multicolumn{1}{c|}{T-BFA} & \multicolumn{1}{c|}{TA-LBF} & TBA (ours) \\ \hline
\multirow{2}{*}{CIFAR-10} & ResNet-18 & \multicolumn{1}{c|}{12.68} & \multicolumn{1}{c|}{673.49} & 16.07 \\ \cline{2-5} 
 & VGG-16 & \multicolumn{1}{c|}{3.43} & \multicolumn{1}{c|}{214.65} & 15.11 \\ \hline
\multirow{2}{*}{ImageNet} & ResNet-34 & \multicolumn{1}{c|}{30.88} & \multicolumn{1}{c|}{205.71} & 25.12 \\ \cline{2-5} 
 & VGG-19 & \multicolumn{1}{c|}{159.86} & \multicolumn{1}{c|}{258.18} & 76.79 \\ \bottomrule
\end{tabular}
}}
\end{table}

\subsection{Statistics of Running Time}
We analyze the running time of the three standard bit flip attacks against quantized models, whose official codes can be accessed. We present the average time used to attack an 8-bit quantized ResNet-18 model with 1,000 different target samples. As shown in Table \ref{tab:runtime}, in CIFAR-10, the heuristic method T-BFA outperforms the two optimization-based methods, TA-LBF and TBA. The running time of T-BFA is highly correlated with the number of bit-flips for it will flip bits one by one until success. For example, attacking VGG-16 utilizing T-BFA costs only 3.43 seconds because it needs only 8.75 bit-flips on average to succeed. For the two optimization-based methods, the time to finish a complete optimization process of TBA is approximately twice that of TA-LBF because the number of parameters involved in TBA is twice that in TA-LBF. However, TA-LBF has to determine suitable hyperparameters in the manner of grid search, making it more time-consuming. For ImageNet, it is usually required to flip more bits to succeed, and our TBA performs better than the other two methods in time efficiency, which can further demonstrate its threat in more complicated tasks. Note that attacking ResNet-34 is more costly than attacking VGG-19 because VGG-19 has a larger number of target parameters as shown in Table \ref{tab:modelinfo}.

\subsection{Training-assisted Baselines}
In the main experiments, we compared only to deployment-only BFAs since training-assisted extension is one of our core contributions. %
However, we also consider comparing our TBA to the training-assisted variants of T-BFA and TA-LBF (FT and FSA cannot be extended) on CIFAR-10 with 8-bit quantized VGG. As shown in Table~\ref{tab:t-baselines}, TBA is on par with or even better than all training-assisted baselines on all metrics.

\begin{table}[t]
\caption{Comparison to the training-assisted variants of baselines. Data points marked in red denote a relatively worse performance.}
\centering
\label{tab:t-baselines}
\scalebox{0.9}{\resizebox{0.5\textwidth}{!}{
\begin{tabular}[\linewidth]{c|ccc:cc:c}
\toprule
Method & ASR (\%) & $N_{flip}$-r & ACC ($M_r$) & $N_{flip}$-f & ACC ($M_f$) & \multicolumn{1}{l}{Time (s)} \\
\hline
T-BFA & 100 & 7.75 & \textcolor{red}{90.73} & 1.01 & \textcolor{red}{87.84} & 38.14 \\
TA-LBF & \textcolor{red}{78.3} & 7.67 & 92.66 & 1.15 & 89.23 & 59.89 \\
TA-LBF-GS & \textcolor{red}{97} & 10.33 & 92.93 & 1.01 & 90.53 & \textcolor{red}{545.26} \\
Ours & 100 & 11.25 & 92.43 & 1.04 & 89.03 & 39.02 \\
\bottomrule
\end{tabular}
}}
\end{table}

\section{Discussions About the Threat Model}
 Our approach differs from the previous BFAs in that we assume the adversary has the access to the training stage and further has the ability to decide the model to be released, which provides a valid reason for the white-box setting generally postulated but without detailed explanation in deployment-time bit flip attacks. In prior BFAs, third-party adversaries usually utilize white-box information like gradients to search for critical bits of the target model's parameters to inject malicious functionality. 
 
 We assume the adversary can implement such a training-assisted attack in at least two cases: (1) The adversary is an insider of one development project, who is in nature able to manipulate the training stage; (2) Utilizing outsourced models is a common phenomenon in the domain of deep learning. In this case, similar to the scenario of backdoor attacks, the adversary can act as an outsider, who releases a high-risk model $M_r$ to the Internet and waits for the victim users to download and then deploy it.

\end{document}